\documentclass[a4paper,fleqn,usenatbib]{mnras}
\usepackage{amsmath}
\usepackage{graphicx}
\usepackage{epstopdf}
\usepackage{threeparttable}
\usepackage{bm}
\usepackage{soul}
\usepackage{amssymb}
\usepackage{hyperref}
\usepackage{blindtext}

\newcommand{\eqb}{\begin{eqnarray}}
\newcommand{\eqe}{\end{eqnarray}}
\newcommand{\ergs}{erg~s$^{-1}$}

\title[Environment and properties of high redshift AGNs]
{Host galaxy properties and environment of obscured and unobscured X-ray selected Active Galactic Nuclei in the COSMOS survey}
\author[Bornancini et al.]{C. Bornancini$^{1,2}$\thanks{bornancini@oac.unc.edu.ar}, D. Garc\'ia Lambas$^{1,2}$\\
$^1$Instituto de Astronom\'{\i}a Te\'orica y Experimental, (IATE, CONICET-UNC), C\'ordoba, Argentina \\
$^2$Observatorio Astron\'omico, Universidad Nacional de C\'ordoba, Laprida 854, X5000BGR, 
C\'ordoba, Argentina\\
}

\begin{document}

\date{Received.../Accepted...}

\pagerange{\pageref{firstpage}--\pageref{lastpage}} \pubyear{2019}

\maketitle

\begin{abstract}
 
We analyse different photometric and spectroscopic properties 
of active galactic nuclei (AGNs) and quasars (QSOs) selected by their mid-IR power-law and X-ray emission 
from the COSMOS survey. We use a set of star-forming galaxies as a control sample to compare with the results. 
We have considered samples of obscured ($HR>-$0.2) and unobscured ($HR<-$0.2) sources including AGNs 
with $L_X<10^{44}$ \ergs, as well as  QSOs ($L_X>10^{44}$ \ergs) with $1.4 \leqslant z \leqslant2.5$.
We also study the typical environment of these samples, by assessing neighbouring galaxy number density and 
neighbour properties such as colour, stellar mass and star formation rate.

We find that the UV/optical and mid-infrared colour distribution of the different AGN types differ significantly. 
Also, we obtain most of AGNs and QSOs to be more compact when compared to the sample of SF galaxies.
	In general we find that the stellar mass distribution of the different AGN sample are similar, obtaining only a difference of $\Delta\overline{\mathrm{log}M}=0.3$ dex ($M_{\odot}$) between unobscured and obscured QSOs.

Obscured and unobscured AGNs and QSOs reside in different local environment at small ($r_p < 100$ kpc) scales.
Our results support previous findings where AGN type correlates with environment.
These differences and those found in AGN host properties cast out the simplest unified model in which obscuration is purely an orientation effect.  

\end{abstract}


\section{Introduction}

Active  galactic  nuclei  (AGN)  are  compact  cores  of some galaxies that emit a high  luminosity  and a non-stellar radiation origin.
The study of AGN offers a great opportunity to know about the nature and evolution of galaxies \citep{miley, hickox09, caplar1, wang17,hickox18, caplar2}. 
It is well known and accepted the model where AGNs are closely related to the presence of a supermassive black hole (SMBH) that accretes the surrounding matter. \citep{zeldo, salpeter, lynden,shakura, rees}. 
After the discovery of quasars (QSOs, \citealt{sch}), we have learned a lot about the nature of AGNs \citep{50years, miley}. 
After these first studies it was thought that AGNs were a rare and a particular case. Today we know, after more than 50 years of research, that, possibly all galaxies are thought to undergo active phases where the central supermassive black hole (SMBH) accretes galactic material \citep{volonteri, catta05, springela,springelb,hop, hop06, hop08}. 

There is also evidence that links some galaxy properties with their presence of a supermassive black hole in their centres. For example, there is amount of observational data that relates black hole and galaxy bulge mass \citep{kor88a, kor88b, mago, marconi03, green, shulze, sani, beifiori, kor13, san14, savor, yang19, shankar} or velocity dispersion \citep{ferra,tremaine,geb,shields,woo,treu,shen,graham, bennert, shen15,saglia}. 

According to their optical spectra, certain AGN types are classified according to the presence of broad emission lines (FWHM $> 2000 \rm \, km \, s^{-1}$, Broad Line AGNs or Type 1) and narrow emission lines or by the absence of broad permitted emission lines (Narrow Line AGNs or Type 2).
In the 1990s a unified model was developed to explain the different observational AGNs features according to the presence of broad or narrow spectral lines \citep{anto,urry,netzer}.
The unified model explains the wide variety of AGN features in terms of the anisotropic geometry of the black hole's immediate surroundings. In this model, AGNs have a central supermassive black hole surrounded by a gaseous accretion disk.  
The broad lines observed in the spectrum would originate in high-speed gas clouds near the SMBH, known as the ``broad line region". 
Beyond this, there would be a torus-shaped structure with sizes of a few parsecs, formed by an optically thick clouds of dust and molecular gas. 
Further away, there would be a structure formed by low density clouds, with dimensions of 100$-$300 pc that move at low speeds producing narrow lines observed in some spectra (known as ``narrow line region"). In this model the observed differences between Type 1 and Type 2 objects are due to orientation effects with respect to the line of sight to the observer. If the torus is face on, it is possible detect the broad-line region directly and the galaxies are identified as Type 1 AGN, otherwise, AGN observed through a torus will be classified as Type 2 AGN.

Originally, this model emerged to explain the differences observed in Seyfert galaxies, that represent a particular case of low luminosity and low redshift AGNs \citep{anto}.
Other unified models include also other types of active galaxies, such as radio galaxies \citep{urry, kembha}. But it is well known that the powerful radio galaxies are a particular case and they are commonly identified with massive elliptical galaxies at low \citep{best,bornan10} redshifts and with massive galaxy systems (such as protoclusters) at high redshifts \citep{vene02, miley04, miley06, hatch11, over16, shima18}. 

There are also other works that attempt some changes to the model. 
\citet{gould12} analysed a sample of local heavily obscured AGNs, i.e., Compton-thick sources with $N_H>1.5\times10^{24}$ cm$^{-2}$, and proposed that the dominant contribution to the observed mid-IR dust extinction is dust located in the host galaxy at much larger scales than predicted for a torus.
These ideas are shared by other authors \citep{malkan,matt,guana,reuna,gould09}, while other authors propose different geometries for
the dusty absorber \citep{nenkova02,elitzur,alonso11,ramos11,audibert}.

If the unified model is correct, that is, the differences observed in the spectra are due only to orientation effects, 
one would expect no differences in host galaxy properties and they would have expected to have similar local galaxy environment, 
but the results obtained at low and high redshifts are contradictory. 
For a sample of low redshift AGNs, \citet{jiang} found differences in halo scale environments between different AGN types.
These authors studied the clustering properties of Type 1 and Type 2 AGNs selected from the Sloan Digital Sky Survey (SDSS), matched in redshift, M$_r$, and $L_{OIII}$, and found similar environments on larger scales. But at scales smaller than 100 $h^{-1}$ kpc, Type 2 AGNs have significantly more neighbours than Type 1 AGNs. These authors concluded that both AGNs samples are hosted by haloes of similar masses, but Type 2 AGNs have more satellites around them.  

Using a sample of low redshift AGNs taken from the SDSS, \citet{xu} calculated the spatial correlation function of galaxies with
Narrow Line Seyfert 1 (NLS1) and Broad Line Seyfert 1 (BLS1) AGNs. These authors found no significant difference
at scales from a few tens of kpc to a few tens of Mpc.

\citet{alle14} investigated the clustering properties of a sample of z$\sim$3 Type 1 and 2 X-ray selected AGNs. 
Their results show that unobscured (Type 1) AGNs inhabit higher-mass haloes compared to the
obscured population (Type 2). They inferred that unobscured AGN reside in 10 times more massive haloes compared to obscured AGN.
Other works also find differences in the dark matter halo masses of different AGN types.
\citet{hickox11} present the first measurement of the clustering
of mid-IR selected obscured and unobscured quasars selected from the 9 deg$^2$ Bo\"{o}tes multiwavelength survey with redshifts in the range $0.7 < z < 1.8$.
These authors found that unobscured and obscured AGNs are located in dark matter halo masses of log($M_{halo}$[h$^{-1}$ M$\odot$])$=12.7$ and 13.3, respectively.
Similar results were found by \citet{donoso} and \citet{dipompeo} for a large sample of AGNs selected from their mid-IR colour using data from the SDSS and the 
Wide-field Infrared Survey Explorer ({\it WISE}) catalogues. \citet{donoso} found that obscured and unobscured AGNs inhabit environments with a typical dark matter halo mass of log($M_h$/$M_{\odot}$ $h^{-1}$) $\sim$13.5 and log($M_h$/$M_{\odot}$ $h^{-1}$)$\sim12.4$, respectively. In a similar way \citet{dipompeo} using a very similar galaxy sample to that of \citet{donoso}, but applying a robust and conservative mask to WISE-selected AGNs, found that obscured quasars reside in haloes of higher mass: log($M_h$/$M_{\odot}$ $h^{-1}$)$\sim13.3$, while unobscured quasar host have log($M_h$/$M_{\odot}$ $h^{-1}$)$\sim12.8$.

In the last years, based on data observations and other results obtained from
numerical simulations, it has taken great relevance evolutionary models that predict 
the existence of stages of formation of galaxies where the AGN would be obscured and
then another period where the AGN would be observed with the appearance of a bright QSO.
One of the pioneering works in this topic was that of \citet{sanders88}.
These authors presented a model for the formation of ultraluminous
infrared galaxies through the strong interaction, or merger
of two  molecular gas-rich spirals.
It was proposed that these ultraluminous infrared galaxies represent the initial, dust-enshrouded stages of
quasars. Once the combined forces of AGN radiation pressure and supernovae explosions
begin to sweep the dust of the nuclear region these
objects can take the appearance of optical quasars. Similar results were also found in numerical simulations \citep{hop,springela, dimatteo}

In \citet{bornan17} we carried out a study using a sample of mid-infrared AGNs selected from the Extended Chandra Deep Field – South (ECDF-S), with spectroscopic and/or photometric redshifts, in addition to information in the X-rays. In \citet{bornan18} we studied host galaxy properties and environment of a sample of Type 1 and 2 AGN taken from the COSMOS2015 catalogue, within 0.3 $\leq z \leq$ 1.1 selected for their emission in X-rays, optical spectra and spectral energy distribution (SED) signatures. In these works we find that obscured AGNs are located in denser local galaxy environments compared to the unobscured AGN sample. 

In this paper we present an analysis of host galaxy properties and environment using an homogeneous and larger AGN sample at high redshifts selected by their mid-IR power-law emission, combining X-ray, optical and spectroscopic observations. We also compare our results with a sample of star-forming galaxies selected with similar properties to the AGN sample.

This paper is organised as follows: In Section 2 we present datasets and samples selection, while in Section 3 we investigate about the properties of obscured and unobscured AGN and QSO host galaxies. The environment of the different AGN and QSO types is analysed in Section 4. In Section 5 we analyse the properties of neighbouring galaxies in the field of AGN and QSO samples. Finally, the summary and conclusions of our work are presented in Section 6. 
Throughout the work we will use the AB magnitude system \citep{oke} and we will assume the same $\Lambda$CDM cosmology adopted by \citet{laigle} with H$_{0} = 70$ km s$^{-1}$ Mpc$^{-1}$, $\Omega_{M} = 0.3$, $\Omega_{\Lambda} = 0.7$.

\section{Datasets}

Observational data used in this paper were obtained from the  Cosmic Evolution Survey (COSMOS, \citealt{scoville}), especially from the COSMOS2015\footnote{The catalogue can be downloaded from ftp://ftp.iap.fr/pub/from\_users/hjmcc/COSMOS2015/} catalogue \citep{laigle}. The COSMOS survey is a multiwavelength photometric and spectroscopic survey 
of 2 deg$^2$, with imaging data from X-ray to radio wavelengths.

The COSMOS2015 catalogue includes GALEX NUV (0.23 $\mu$m) observations \citep{zamo}, CFHT/MegaCam (u$^*$-band) \citep{sanders07}, Subaru Suprime-Cam optical bands (B, V, g, r, i, z$^{++}$) \citep{tani07,tani15}. Also near-infrared YJHKs-band data taken with WIRCam and Ultra-VISTA data \citep{mccrac10,mccrac12}. Four Spitzer Infrared Array Camera (IRAC) bands ($[3.6], [4.5], [5.8]$ and $[8.0]$ $\mu$m), S-COSMOS \citep{sanders07}, the Spitzer Extended Mission Deep Survey \citep{ashby13} and the Spitzer S-CANDELS survey \citep{ashby}. 
In addition, there are several X-ray surveys in this area: $XMM$-COSMOS \citep{hasi07,cappe}, C-COSMOS \citep{elvis,civano16} and the NuSTAR survey \citep{civano15}. 

\subsection{AGN and SF galaxy sample selection}
\label{agn}

\begin{figure}
 \centering 
\includegraphics[width=0.47\textwidth]{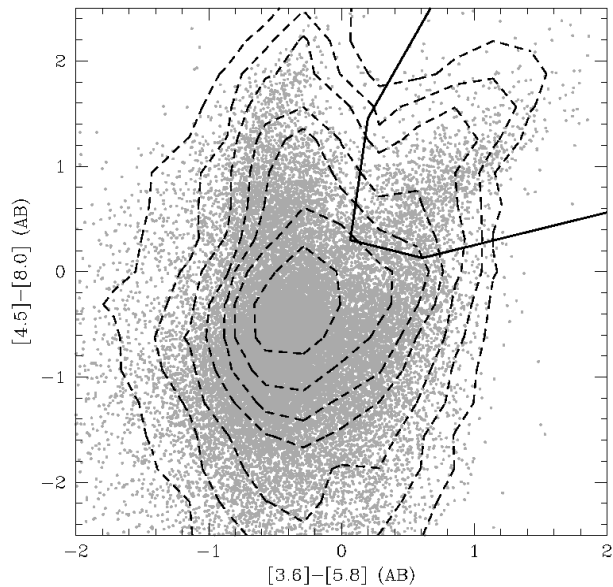}
	\caption{$[4.5]-[8.0]$ vs. $[3.6]-[5.8]$ colour-colour diagram for objects selected in 24 $\mu$m.
	The area enclosed by the thick solid lines represents the selection criteria of \citet{chang} for AGN candidates with power law emission and 
	flux densities monotonically rising from 3.6 to 8 $\mu$m.}
\label{power}
\end{figure}

The main goal of our work is to analyse the environment and properties of X-ray selected AGNs and the galaxies located in their environments.
For this, we selected a sample of AGNs along with a comparison sample formed by SF galaxies with $1.4 \leqslant z \leqslant2.5$.
The sample of AGNs used in this work was selected from the \citet{chang} catalogue. 
This catalogue is based on objects with mid-IR (MIR) 24 $\mu$m emission ($S_{24\mu m}$ $\geq$ 80$\mu$Jy, \citealt{lefloch09}). 
These authors matched this sources with the COSMOS2015 catalogue \citep{laigle} in order to obtain photometry from the optical to the far-IR (FIR).
According to a colour–colour selection, they have selected IR power-law AGN candidates \citep{lacy04}.
It is well known that galaxies dominated by AGN emission typically show a power law SED at near- and mid-IR wavelengths.
In Figure \ref{power} it can be seen the $[4.5]-[8.0]$ vs. $[3.6]-[5.8]$ colour-colour diagram for objects detected in 24 $\mu$m. 
The box represents the selection criteria presented by \citet{chang} for AGN candidates with power law emission with flux densities monotonically rising from 3.6 to 8 $\mu$m.

The derived catalogue was then cross-matched with the X-ray catalogue presented by \citet{marchesi}. These authors identified 1770 X-ray sources from the master spectroscopic catalogue available for the COSMOS collaboration (M. Salvato et al., in preparation).

The selection of AGNs with X-ray emission presents great advantages to those selected in the optical or in the IR. 
X-ray selected AGNs are less affected by obscuration and contamination from
non-nuclear emission, mainly due to star formation processes, which are far less significant than in optical and infrared surveys \citep{donley08,donley12,lehmer12,stern12}.

In this paper we study a sample of obscured (absorbed) and unobscured (unabsorbed) AGNs and QSOs in the X-rays. 

To accomplish it, we analyse the distribution of the AGNs in the hardness ratio vs. hard-X-ray luminosity diagram.
The hardness ratio (HR, which is an indication of the X-ray spectral shape) of each X-ray source using the following relation,
\begin{equation}
{\rm HR}=\frac{H-S}{H+S},
\end{equation}
where $H$ and $S$ are the count rates in the hard ($2-10$ keV) and soft ($0.5-2$ keV) bands, respectively. 

\begin{figure}
 \centering 
\includegraphics[width=0.47\textwidth]{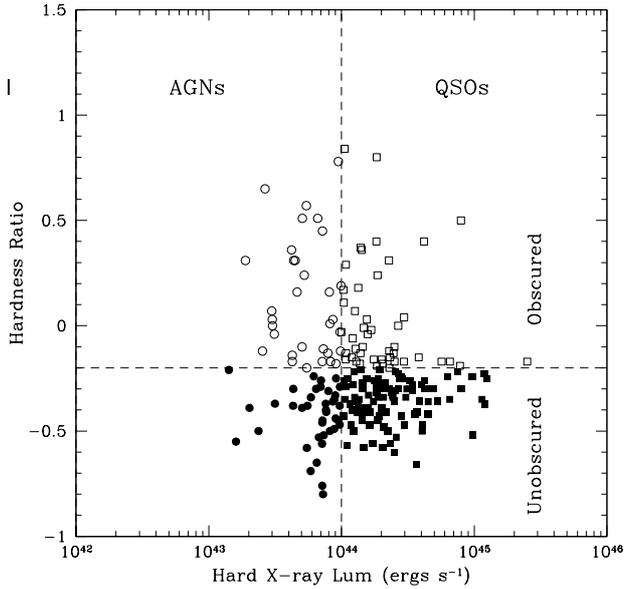}
	\caption{Hardness ratio as a function of hard ($2-10$ keV) X-ray luminosity.
	Open and filled circles and squares represent obscured and unobscured AGNs and QSOs, respectively.
	Vertical dashed lines show the typical separation for AGNs and QSOs used in the X-rays.
The dashed horizontal line shows the HR value for a source with $N_H>10^{21.6}$ cm$^2$ at $z>1$ which is used to separate obscured and unobscured sources in the X-rays. }		
\label{HR}
\end{figure}

\begin{figure*}
 \centering 
\includegraphics[width=0.47\textwidth]{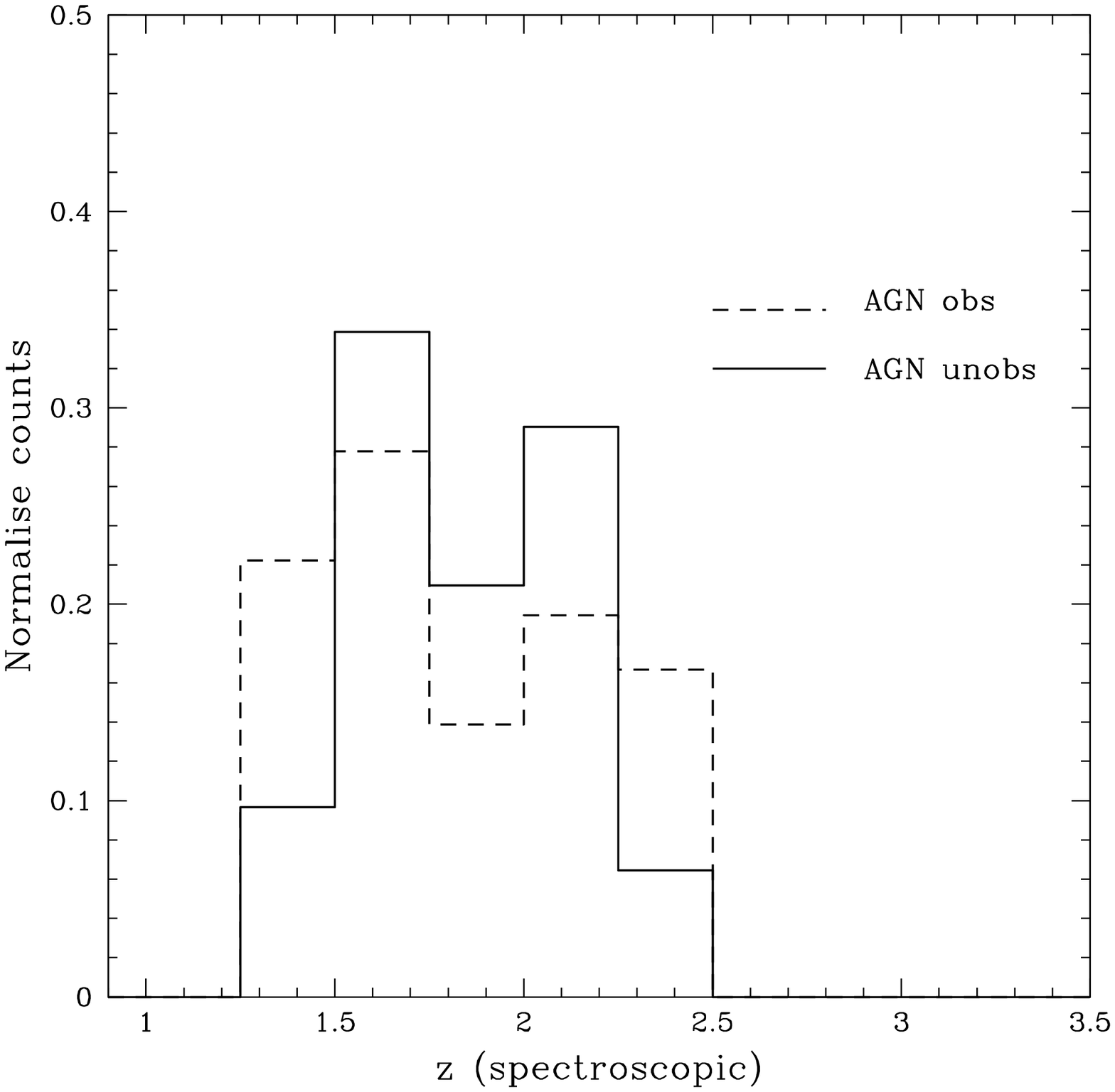}
\includegraphics[width=0.47\textwidth]{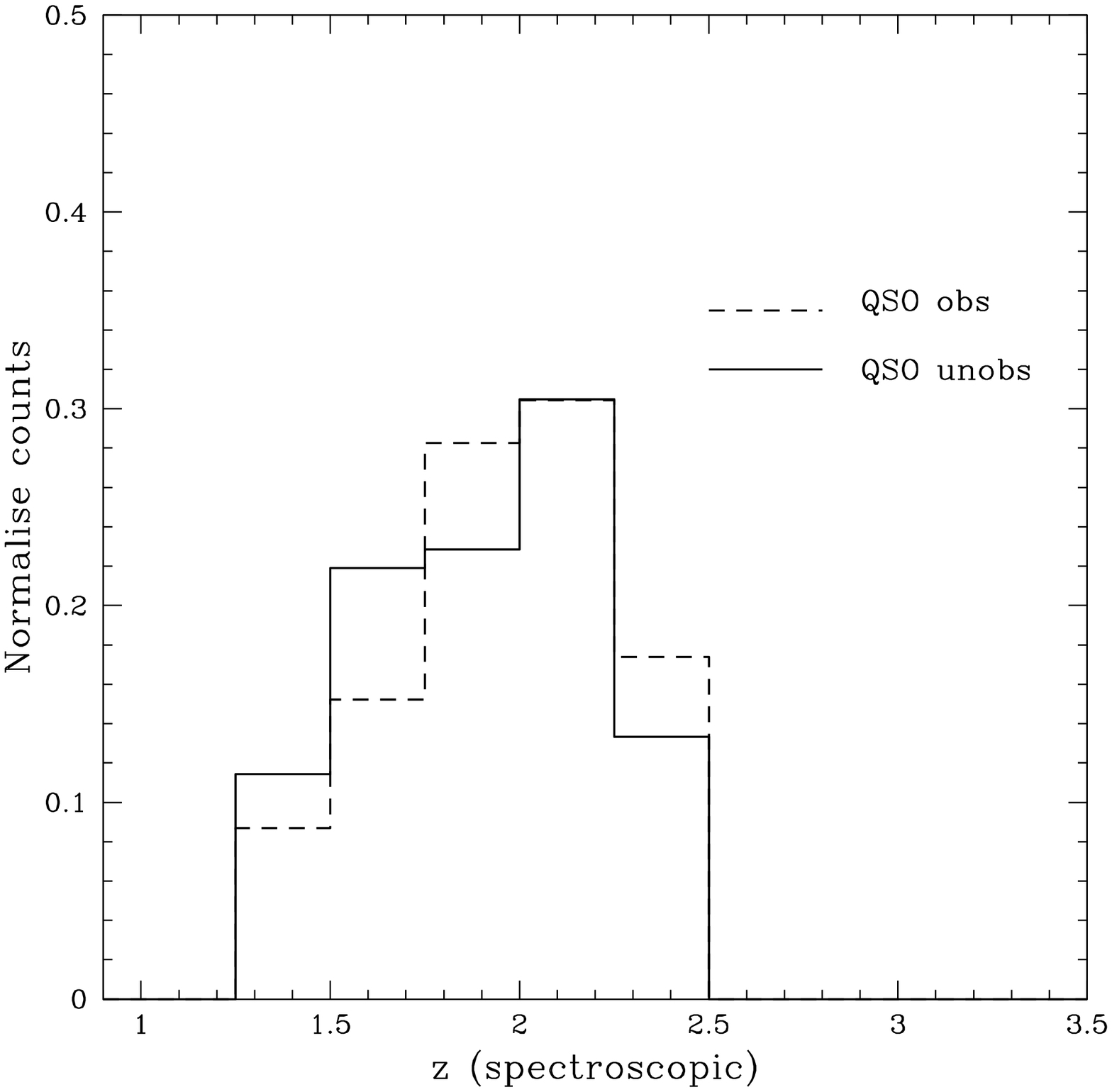}
	\caption{Normalised redshift distribution for the final selected obscured (dashed line histogram) and unobscured (solid line histogram) AGNs (left panel) and the corresponding distribution for QSOs (right panel). }
\label{zspec}
\end{figure*}

In Figure \ref{HR} we plot the hardness ratio as a function of hard X-ray luminosity ($2-10$ keV) for a sample of
AGNs with $1.4 \leqslant z_{spec} \leqslant2.5$.
The dashed horizontal line shows the HR value (HR$=-0.2$) for a source with a neutral hydrogen column density, $N_H>10^{21.6}$ cm$^2$ at $z>1$ \citep{gilli09}, which is used by several authors \citep{gilli09, treis09, marchesi} to separate obscured and unobscured sources in the X-rays.
This is due to the the fact that soft X-ray emission of obscured AGNs tend to be absorbed, while hard X-ray are able to escape.

To select the AGN and QSO samples we use the selection criteria according to the luminosity in the hard X-rays, i.e., sources with hard X-ray luminosities $L_X<10^{44}$ \ergs are classified as AGNs, while sources with  $L_X \ge 10^{44}$ \ergs are classified as QSOs \citep{treis09}.

The final AGN sample consists of 36 and 62 obscured and unobscured AGNs and 46 and 105 obscured and unobscured QSOs.
In Figure \ref{zspec} we plot the spectroscopic redshift distribution for the final sample of AGNs and QSOs (obscured and unobscured, respectively).
We applied a Kolmogorov-Smirnov (KS) test on the redshift distribution of both AGN and QSO samples. We find that the redshift distribution are similar at 72\% and 65\% for the sample of obscured and unobscured QSO and AGN samples.

In order to compare some properties of the AGNs sample, we have selected a sample of SF galaxies. 
For this reason we use two different selection criteria. First, we used a Lyman-break selection technique identified in the $NUV-B$ versus $B-V$ diagram proposed by \citet{ly09} to select SF galaxies with z$\approx$1.8$-$2.8. In Figure 4, left panel, we plot $NUV-B$ vs. $B-V$ colours of galaxies with $B< 27$, which is the 3$\sigma$ detection limit \citep{laigle}. The box represents the empirical selection criteria for the LBG sample with z$\approx$1.8$-$2.8 proposed by \citet{ly09}.

\begin{figure*}
\includegraphics[width=80mm]{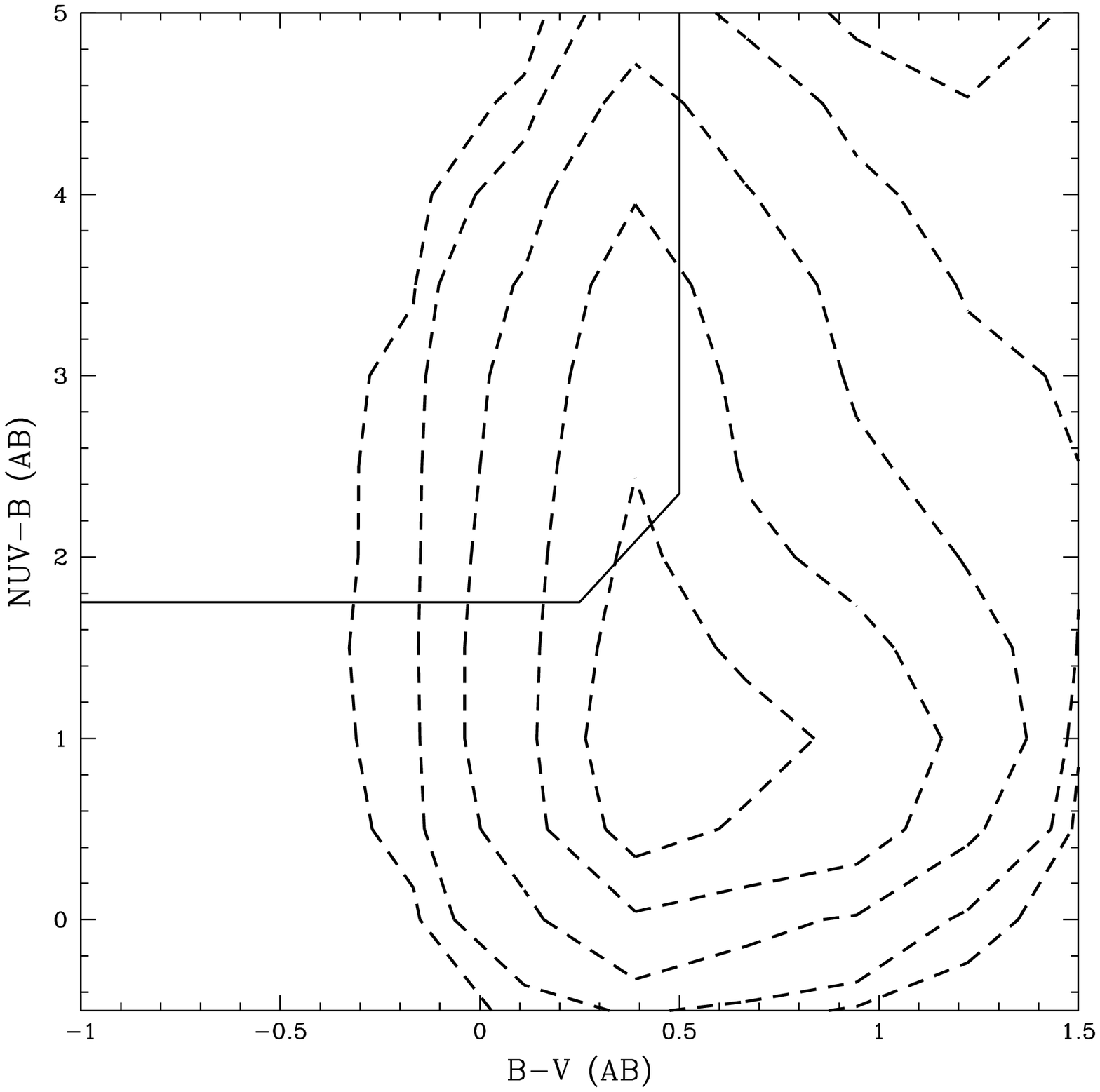}
\includegraphics[width=80mm]{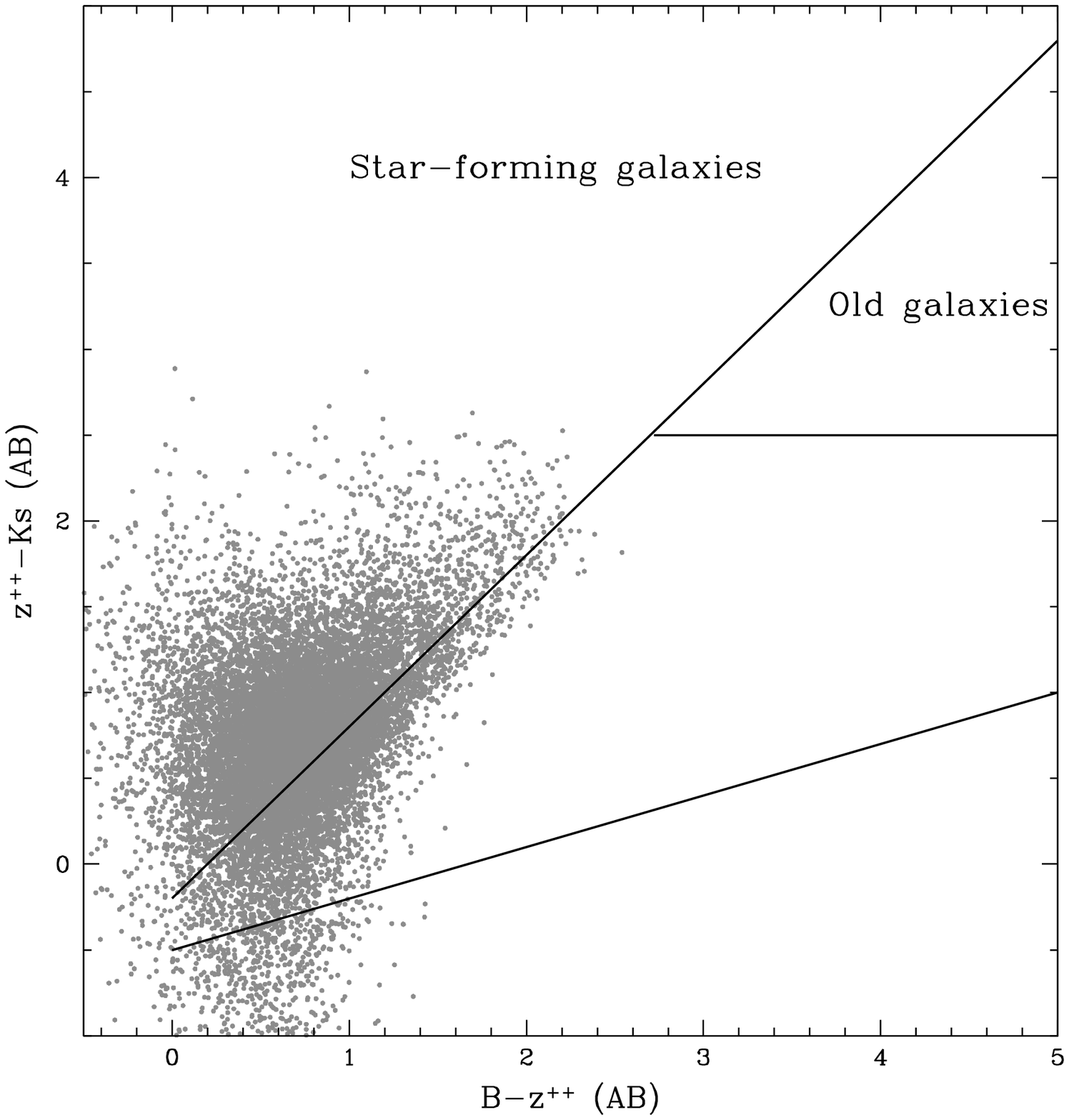}
\label{sf}
	\caption{Left panel: $NUV-B$ vs. $B-V$ colour of galaxies with $B< 27$. The upper left box represents the empirical selection criteria for LBG galaxies proposed by \citet{ly09}. Right panel: $z'-K$ vs. $B-z'$ diagram illustrating the selection criteria for the BzK samples, showing the sBzK (star-forming galaxies) and pBzK (passive old galaxies) regions. Gray points represents objects previously selected with the Lyman-break selection technique shown in the left panel.}
\end{figure*}

In order to avoid contamination by possible objects that are not associated with SF galaxies, we used the BzK selection technique \citep{daddi,kong}. This photometric technique was designed to identify star-forming (sBzK) and passive (pBzK) galaxies at $z > 1.4$.
In Figure 4, right panel, we plot $z'-K$ vs. $B-z'$ colour-colour diagram with the delimited regions occupied by star-forming and passive old galaxies \citep{daddi}. We plot objects previously selected with the Lyman-break selection technique identified in the $NUV-B$ vs. $B-V$ diagram. As it can be seen, some objects have colours outside the selection region of SF galaxies. For this reason we have excluded this objects from our SF galaxy sample.

We find that only a small percentage ($<$3\%) of SF galaxies have spectroscopic redshift measurements, so we decided to use only photometric redshifts.
In order to select objects with good photometric redshift estimations, we chose those whose width of the 68\% confidence interval is less than 0.1. 
After this, we choose those objects with the same photometric redshift range $1.4\leqslant z_{ph}\leqslant 2.5$ and with $M_K<-22$, whose values match those found in the AGN/QSO samples. The number of SF galaxies obtained is 3157 objects.

\begin{figure*}
 \centering 
\includegraphics[width=0.47\textwidth]{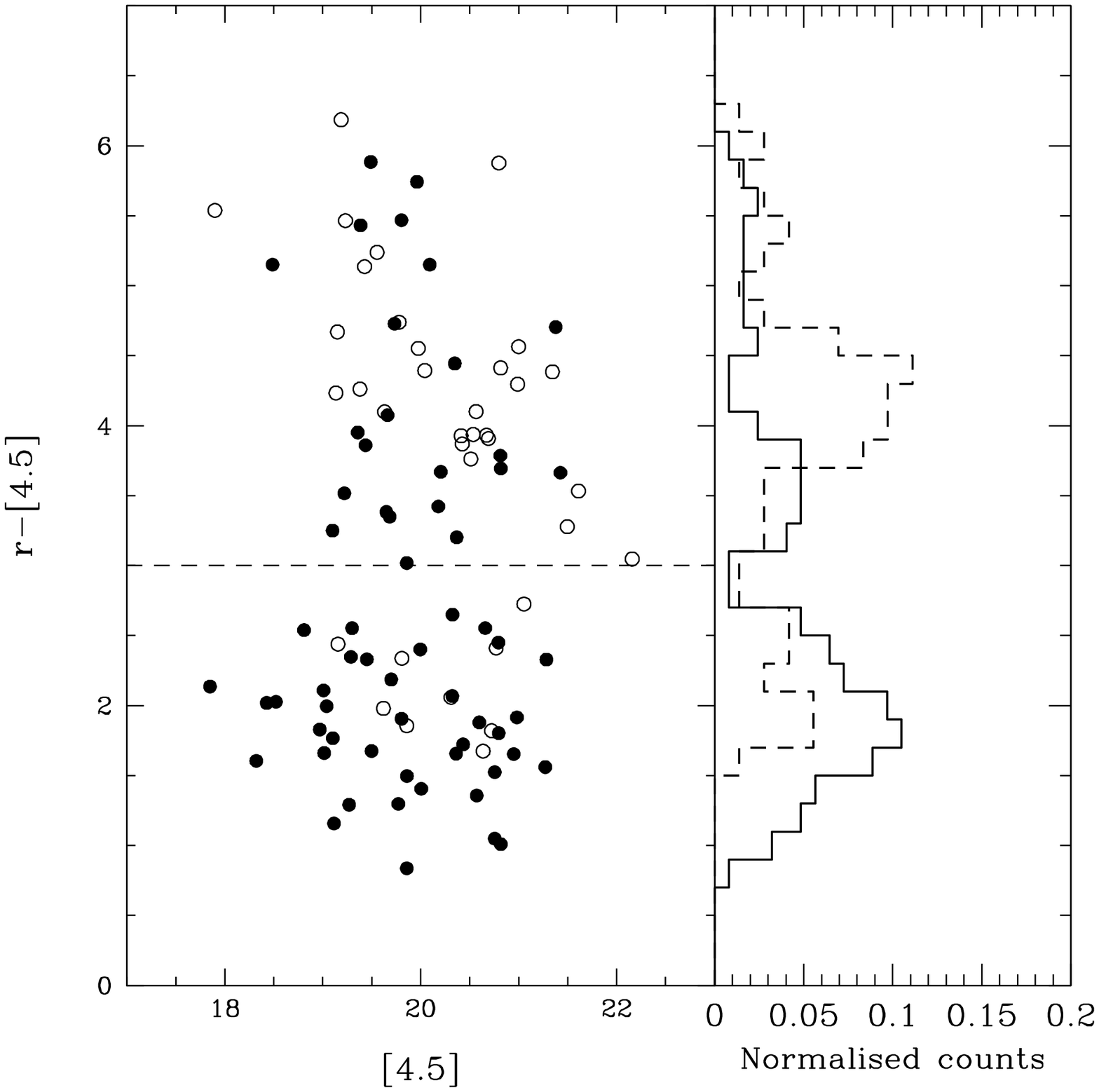}
\includegraphics[width=0.47\textwidth]{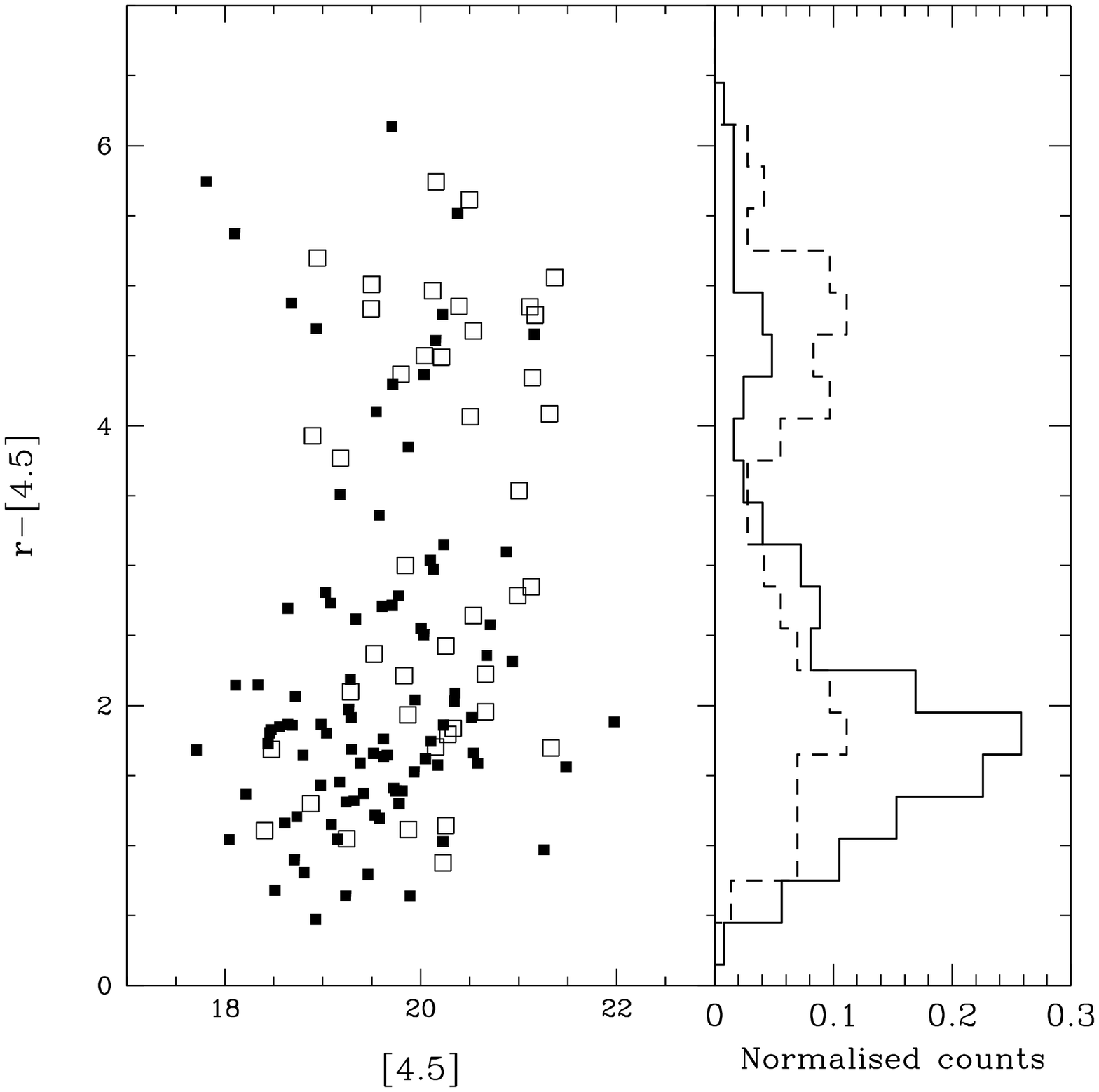}
	\caption{Left: optical-IR colour, $r -[4.5]$ vs. 4.5 $\mu$m for obscured (open circles) and unobscured (filled circles) AGNs.
	         Right: the corresponding colours for obscured (open squares) and unobscured (filled squares) QSOs. Right panels in each figure show the corresponding $r -[4.5]$ colour distribution of obscured (dashed lines) and unobscured (solid line histogram) AGN and QSO samples. }		
\label{r45}
\end{figure*}



\begin{figure*}
\includegraphics[width=80mm]{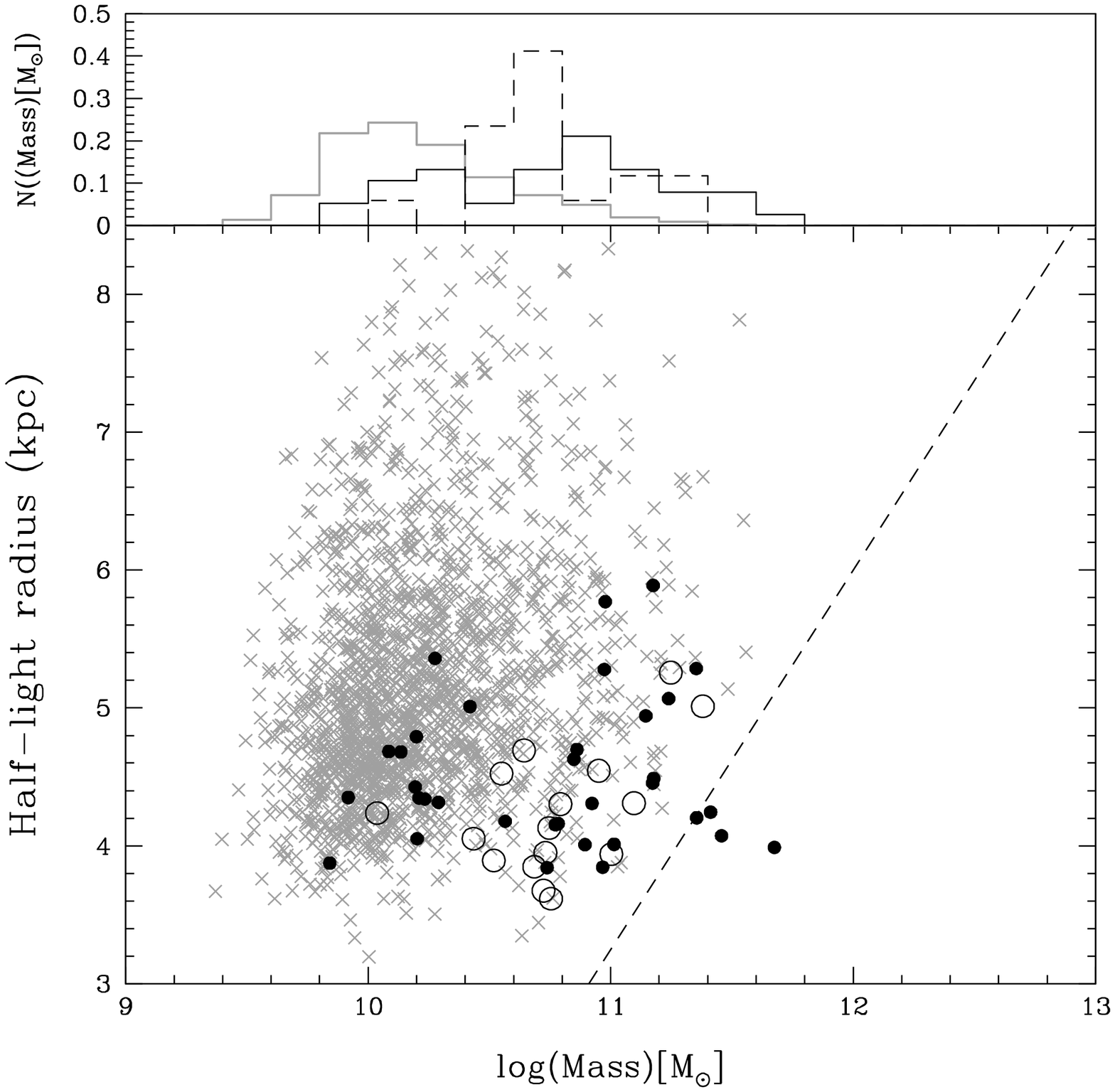}
\includegraphics[width=80mm]{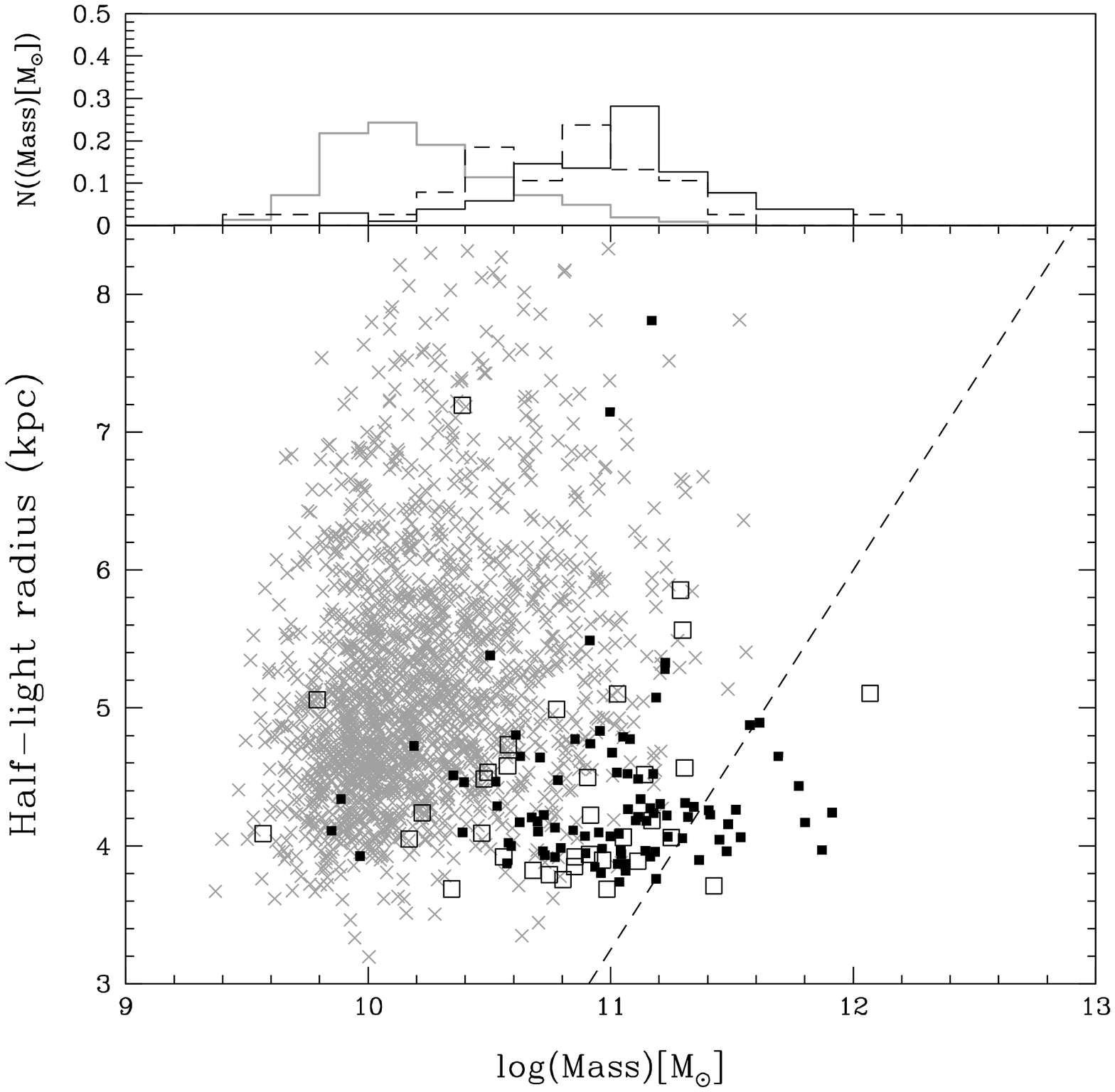}
\caption{Stellar mass$-$size relation for our sample of obscured and unobscured (open and filled circles) AGNs (left panel) and for 
	obscured and unobscured (open and filled squares) QSOs (right panel). Gray crosses represent values obtained for SF galaxies. 
	The dashed black line represent the selection criterion for compact galaxies proposed 
	by \citet{barro13}. The upper panels in each figure show the stellar mass distribution for obscured and unobscured (dashed and solid lines) AGNs (left panel), obscured and unobscured (dashed and solid lines) QSOs (right panel) and SF galaxies (gray dot line). }
\label{masssize}
\end{figure*}

\section{Host galaxy properties}

\subsection{Mid-infrared colours}

In this section we study the UV/optical/IR colour of the AGN and QSO samples.
In Figure \ref{r45} we plot optical-IR colour, $r -[4.5]$ vs. 4.5 $\mu$m for obscured and unobscured AGNs (left panel) and QSOs (right panel).
In each case we also show the corresponding $r -[4.5]$ colour distribution.
As previous works, \citet{hickox07, hickox11, bornan17, bornan18}, we find for the AGN sample that the $r -[4.5]$ colour distribution is bimodal.
For the sample of AGNs we find that the population of obscured AGNs consists mostly of objects with predominantly red colours ($r -[4.5] > 3$), in addition to a small population of AGNs with blue colours (see Figure \ref{r45}, left panel).
For unobscured AGNs we find a similar but in reverse result, i.e., it consists mostly of objects with predominantly blue colours, in addition to a small population of AGNs with red colours (see Figure \ref{r45}, left panel). In the case of obscured QSOs (see Figure \ref{r45}, right panel), we observe two similar populations of objects with red and blue colours, while for the sample of unobscured QSOs, we find that most objects have predominantly blue colours with a peak near $r -[4.5]\sim 1.5$. 

In order to determine and quantify the bimodality found in the optical-IR colours, we have used the GMM code of \citet{muratov}. 
This code performs a Gaussian mixture modeling (GMM) test and uses three indicators to distinguish unimodal and bimodal distributions: the kurtosis of the distribution, the separation of the peaks ($D$), and the probability of obtaining the same $\chi^2$ from a unimodal distribution ($p(\chi^2)$). 
The separation of the peaks ($D$) is calculated as 

\begin{equation}
D=\frac{\big|\mu_1-\mu_2 \big|}{\sqrt{\frac{\sigma^2_1+\sigma^2_2}{2}}}
, 
\end{equation}

where $\mu_1$ and $\mu_2$ are the mean values of the two peaks of the proposed bimodal distribution, and $\sigma_1$, $\sigma_2$ are the corresponding standard deviations. The distribution is considered bimodal by required negative kurtosis, $D > 2$, and $p(\chi^2)<0.001$.

For the sample of obscured AGNs, we find kurtosis$=-0.85$, $\mu_1=2.1\pm0.3$, $\mu_2=4.4\pm0.2$, $D=3.9\pm0.6$ and $p(\chi^2)<0.001$. 
For unobscured AGNs, we find kurtosis$=-0.41$, $\mu_1=2.1\pm0.2$, $\mu_2=4.5\pm0.5$, $D=3.2\pm0.9$ and $p(\chi^2)<0.001$. 
While, for the sample of obscured QSOs, we find kurtosis$=-1.50$, $\mu_1=1.9\pm0.2$, $\mu_2=4.7\pm0.1$, $D=4.6\pm0.6$ and $p(\chi^2)<0.001$ and
 for unobscured QSOs, we find kurtosis$=+0.95$, $\mu_1=1.7\pm0.1$, $\mu_2=4.1\pm0.6$, $D=3.3\pm1.4$ and $p(\chi^2)<0.001$

The colour separation limit matches approximately the value found by \citet{hickox11,bornan17} which corresponds to  $r -[4.5]\sim 3$, for the sample of obscured and unobscured AGNs and for obscured QSOs. For unobscured QSOs we find that $r-[4.5]$ colour distribution it is not bimodal, because kurtosis is positive.

\begin{figure*}
 \centering 
\includegraphics[width=0.49\textwidth]{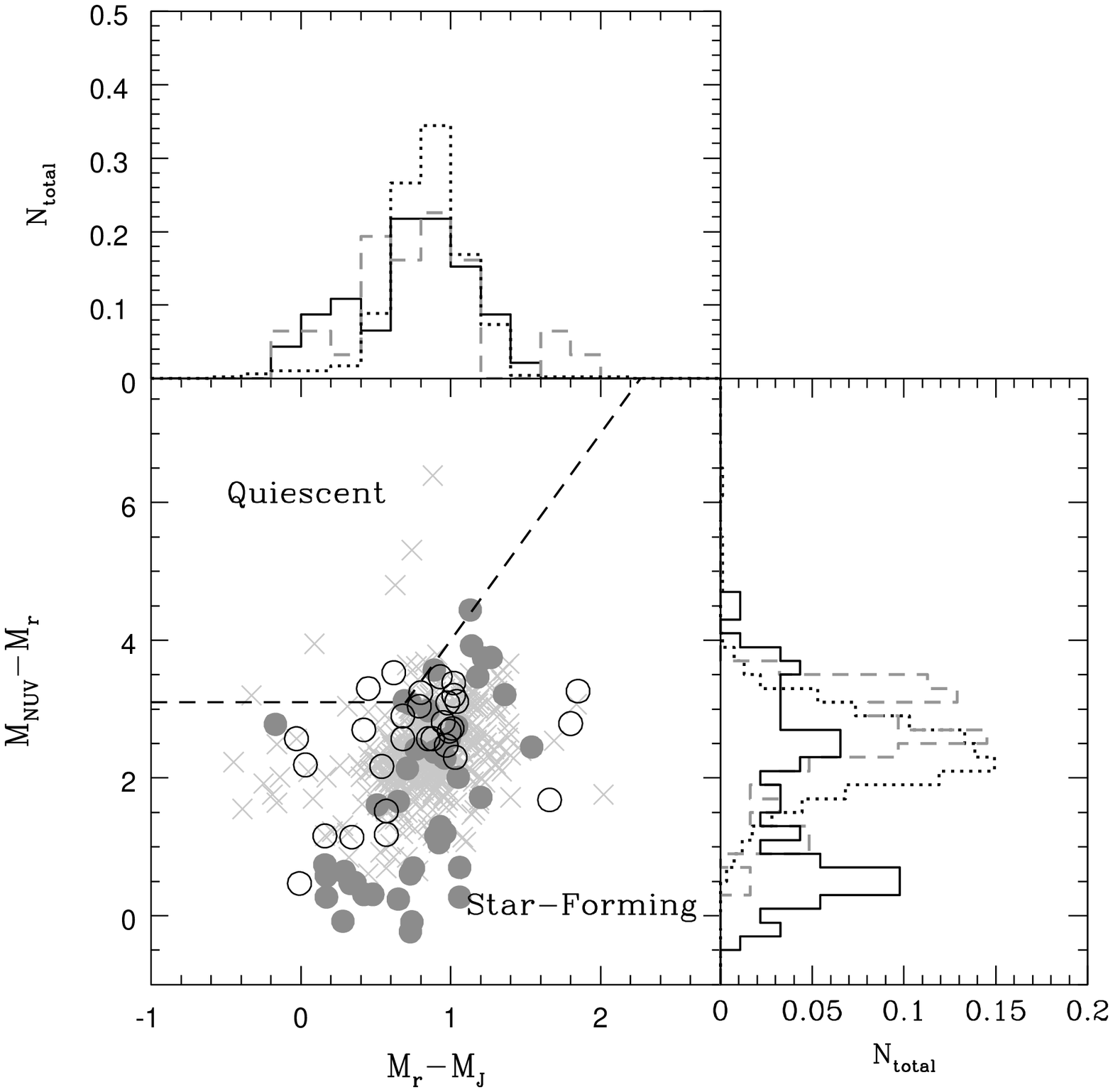}
\includegraphics[width=0.49\textwidth]{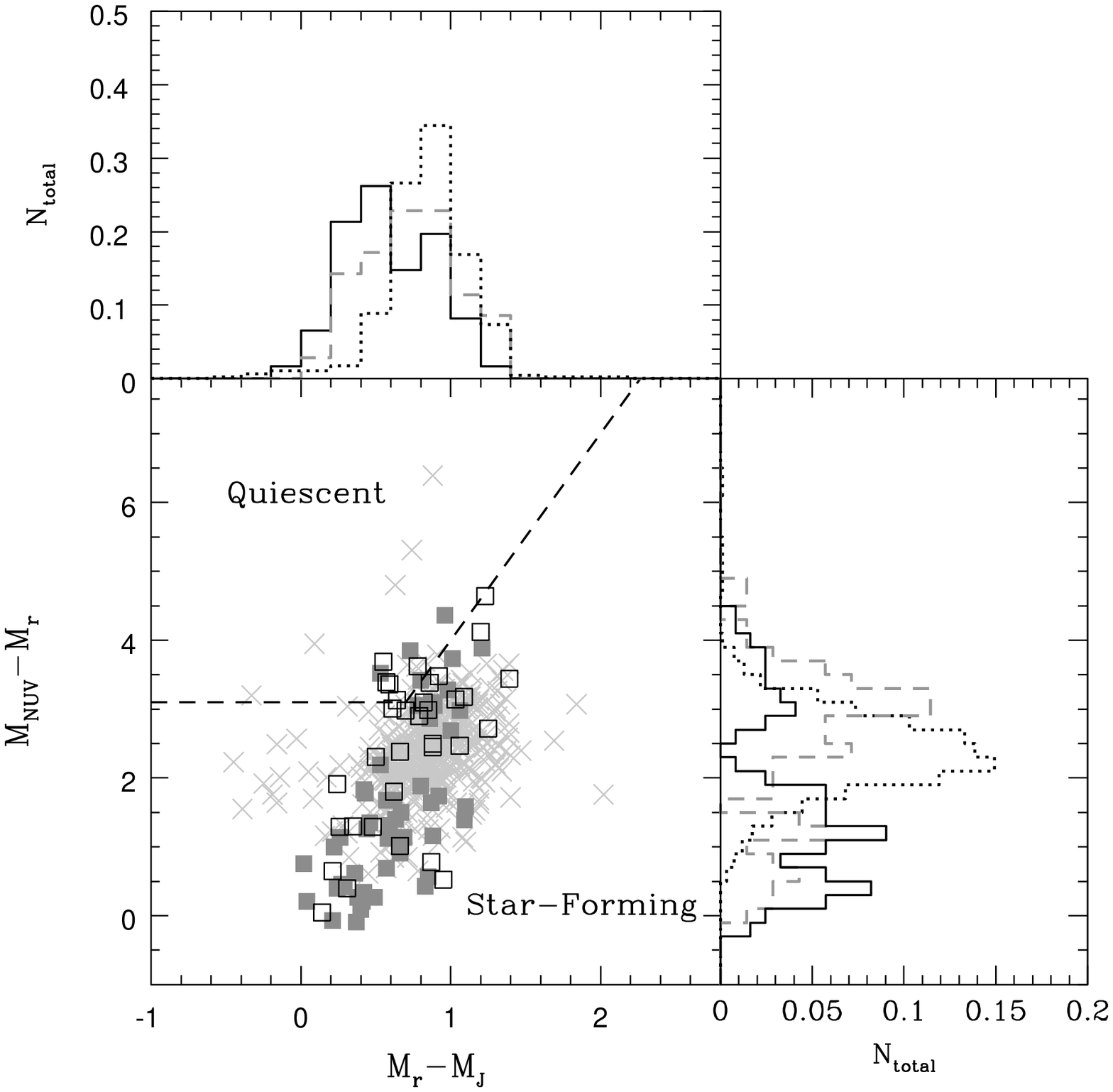}
	\caption{Rest-frame $M_{NUV}-M_r$ vs. $M_r-M_J$ colour diagram. Gray crosses represent SF galaxies, while open and filled circles represent obscured and unobscured AGNs (left panel). Right panel: Same as left panel for the sample of obscured (open squares) and unobscured (filled squares) QSOs. Dashed lines show the limits used to empirically separate quiescent from star-forming galaxies taken from \citet{ilbert13}. Upper and right panels show the $M_r-M_J$ and $M_{NUV}-M_r$ colour distribution for obscured and unobscured AGNs (dashed and solid lines) and for obscured and unobscured QSOs (dashed and solid lines). Dot histograms in upper and right panels show the corresponding colours for the sample of SF galaxies.}

\label{quiescent}
\end{figure*}

\subsection{Mass-size relation}
\label{ms}

In this section we analyse the mass-size relation for the AGN and QSO samples.
In the local Universe it is well known that galaxies are distributed in two large groups: one formed by
passive galaxies with predominantly red colours and old stellar populations, and another group consisting of
blue, star-forming galaxies \citep{kauff03, baldry}.
In addition, these objects have different morphologies. At a given stellar-mass, star-forming 
galaxies are typically larger and have disk-like morphologies, while passive ones are represented by spheroids with large concentrations of light \citep{williams10, wuyts}.

The discovery of passive galaxies at z$\sim$ 2, with small sizes, called compact quiescent galaxies or red nuggets \citep{Dam} showed these objects to be $\sim$5 times smaller than local, equal-mass analogues \citep{daddi05,truji07,toft07,cassata,szo}, in contrast with star-forming galaxies at high redshifts identified with large disks \citep{kriek,schrei}.

For this reason, we cross-correlated our AGN and QSO samples with the catalogue presented in \citet{bongio12}. These authors calculated colours, stellar masses and star-formation rates (SFR) of a sample of X-ray AGNs selected from the $XMM$-COSMOS catalogue. These parameters were carefully calculated using their spectral energy distributions, which have been parametrised using a two-component (AGN$+$galaxy) model fit.
Figure 6 shows a mass-size diagram for our sample of obscured and unobscured (open and filled circles) AGNs (left panel) and for obscured and unobscured (open and filled squares) QSOs (right panel). We also plot the corresponding values for the SF galaxy sample (gray crosses). For all samples we have used the half-light radii taken from the \citet{laigle} catalogue. Dashed lines in Figure \ref{masssize} show the selection criterion for compact galaxies proposed by \citet{barro13}, $M/r_{e}^{1.5} =$ 10.3 $M_{\odot}$ kpc$^{-1.5}$.
As proposed by \citet{barro13,barro14}, the right region encloses most of the quiescent population at z $>$1.4.
In general we find that most of AGNs and QSOs are more compact compared to the sample of SF galaxies.
While only 12\% of unobscured AGNs reside in the region occupied by compact massive galaxies, we find that 15\% and 5\% of the unobscured and obscured QSOs reside in this region.

We have also included the corresponding stellar mass distribution for the samples of AGNs, QSOs and SF galaxies. The stellar mass of the SF galaxy sample were taken from the catalogue of \citet{laigle} (MASS\_BEST).
As it can be seen, AGNs and QSOs are identified with massive host galaxies compared to the sample of SF galaxies. The stellar mass distribution of obscured and unobscured AGNs are similar. For the obscured and unobscured AGNs we find $\overline{\mathrm{log}M_{obs}}=10.7$ ($M_{\odot}$), $\sigma_{obs}=0.3$ and $\overline{\mathrm{log}M_{unobs}}=10.8$ ($M_{\odot}$), $\sigma_{unobs}=0.5$, respectively. For the sample of QSOs, we find a difference of $\Delta\overline{\mathrm{log}M}=0.3\pm 0.1$ dex ($M_{\odot}$), obtaining $\overline{\mathrm{log}M_{obs}}=10.7$ ($M_{\odot}$), $\sigma_{obs}=0.5$ and $\overline{\mathrm{log}M_{unobs}}=11.0$ ($M_{\odot}$), $\sigma_{unobs}=0.5$. For the sample of SF galaxies we find $\overline{\mathrm{log}M_{SF}}=10.2$ ($M_{\odot}$), $\sigma_{obs}=0.3$. This last result is in contradiction with those found by \citet{bongio12} and \citet{bornan18}.
\citet{bongio12} found that the stellar mass distribution of obscured and unobscured AGN is nearly
independent of the AGN type. Analysing a sample of Type 1 and 2 AGNs with $z<$1, \citet{bornan18} found that, despite uncertainties, the mean stellar mass of Type 2 sources is 0.2 dex higher than that of Type 1 (see discussion in \citealt{zou}).

\subsection{UV, optical and near-IR colours}

It is well known that galaxy colours in the near-UV ($NUV$), optical ($r-$band), near-infrared ($J-$band or $K-$band), or with similar filters can be used to separate star-forming from quiescent galaxies \citep{williams,arnouts13,ilbert13,lin16,moutardII}.
One of the first works using this colour-colour diagram was that of \citet{williams}. These authors found
that star-forming and quiescent galaxies occupy two distinct regions in the rest-frame ($U-V$) vs. ($V-J$) colour-colour diagram.
\citet{arnouts13} analysing a sample of galaxies with spectroscopic redshifts between 0.2$<z<$1.3 in the COSMOS survey, found a very good correlation between the specific star-formation (sSFR, which measures the SFR per unit galaxy stellar mass) and the position of these two galaxy populations in the $NUV-r$ vs. $r-K$ diagram. Galaxies located in the quiescent region have low sSFR values such as those found in passive galaxies with little or null SFR, showing that the $NUV-r$ colour is a good tracer of the specific SFR, since the NUV band is sensitive to recent star formation and the $r-$band to old stellar populations.
This method is also valid for separating galaxy morphological types. \citet{patel} found that the quiescent region is dominated by galaxies with strong bulge components while the star-forming region is comprised of disks.

Figure \ref{quiescent} shows the rest-frame $M_{NUV}-M_r$ vs. $M_r-M_J$ colour diagram for obscured and unobscured (open and filled circles, respectively) AGNs (left panel) and for obscured and unobscured (open and filled squares) QSOs (right panel).
The dashed lines representing the boundaries that separate regions occupied by quiescent and star-forming galaxies. Quiescent objects are those with $M_{NUV}-M_r > 3(M_r-M_J)+1$ and $M_{NUV}-M_r > 3.1$ while star-forming galaxies occupy regions outside of this criterion \citep{ilbert10, ilbert13}.
We also included the corresponding colours for SF galaxies. According to the results obtained in the Section \ref{ms}, 
and with the aim at comparing host galaxies with similar stellar masses, we select SF galaxies with log($M_{\odot}$) $>$ 10.5 (gray crosses). The final number of SF galaxies is 462 objects.

As it can be seen, the majority of the AGNs and QSOs sample falls inside the region occupied by the star forming galaxy population. 

In \citet{bornan18} we conducted a similar study for a sample of Type 1 and 2 AGN with 0.3 $\leq z \leq$ 1.1 selected for their emission in X-rays, optical spectra and SED signatures. In this work we found that Type 1 AGNs are exclusively located in the region populated by star-forming galaxies in the $M_{NUV}-M_r$ vs. $M_r-M_J$ colour diagram, whereas Type 2 are divided between the region occupied by quiescent (29\%) and star-forming galaxies (71\%). The majority of the AGN sample analysed in this work have hard X-ray luminosities similarly to those found in normal AGNs or Seyfert-like galaxies ($10^{42}<L_X<10^{44}$ \ergs). Only 7\% and 20\% of Type 2 and 1 objects were found to have hard X-ray luminosities similarly to those in QSOs ($L_X>10^{44}$ \ergs).  

In the present work we also find bimodalities in $M_{NUV}-M_r$ colour distribution in unobscured AGN and QSO samples and in obscured QSOs. 
The results obtained using the GMM code are as follows:
For the obscured QSO sample, we find kurtosis$=-1.52$, $\mu_1=0.3\pm0.1$, $\mu_2=3.0\pm0.3$, $D=4.6\pm1.0$ and $p(\chi^2)<0.001$. 
For the unobscured QSO sample, we find kurtosis$=-0.80$, $\mu_1=1.0\pm0.1$, $\mu_2=3.3\pm0.2$, $D=4.5\pm0.6$ and $p(\chi^2)<0.001$. 
While for unobscured AGN, kurtosis$=-1.15$, $\mu_1=0.6\pm0.1$, $\mu_2=2.7\pm0.3$, $D=3.4\pm0.6$ and $p(\chi^2)=0.002$.
In the case of obscured AGNs, the colour distribution are not bimodal and we have obtained: kurtosis$=+1.3$, $\mu_1=1.9\pm0.9$, $\mu_2=2.9\pm0.8$, $D=1.9\pm2.0$ and $p(\chi^2)=0.03$.

Unobscured AGNs and QSOs have $M_{NUV}-M_r$ colours slightly blue than SF galaxies. We find the following parameters for the SF galaxy sample for a unimodal distribution: kurtosis$=+4.85$, $\mu_1=2.40\pm0.03$, and $\sigma=0.60\pm0.03$.

\begin{figure*}
\centering 
\includegraphics[width=0.47\textwidth]{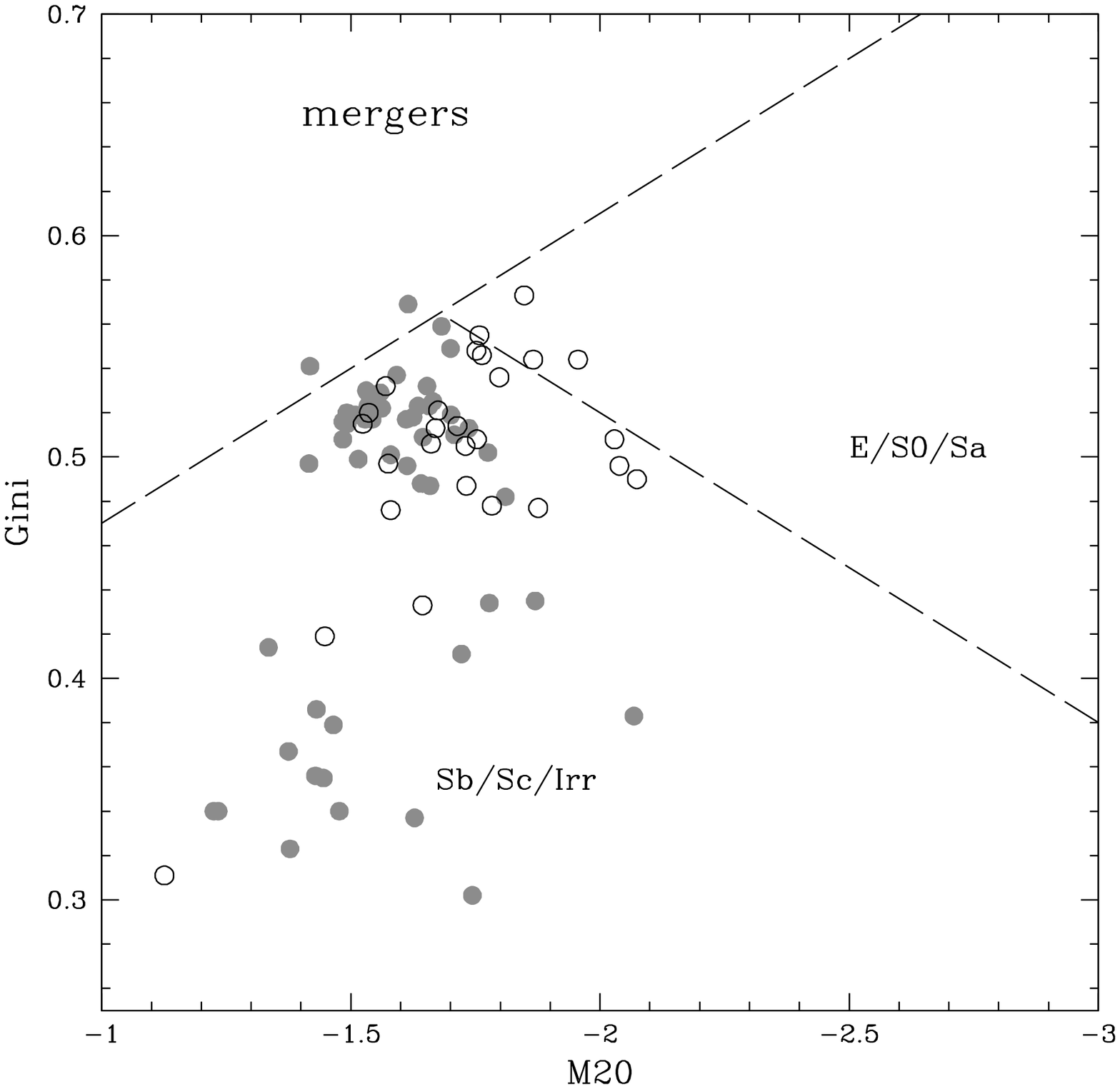}
\includegraphics[width=0.47\textwidth]{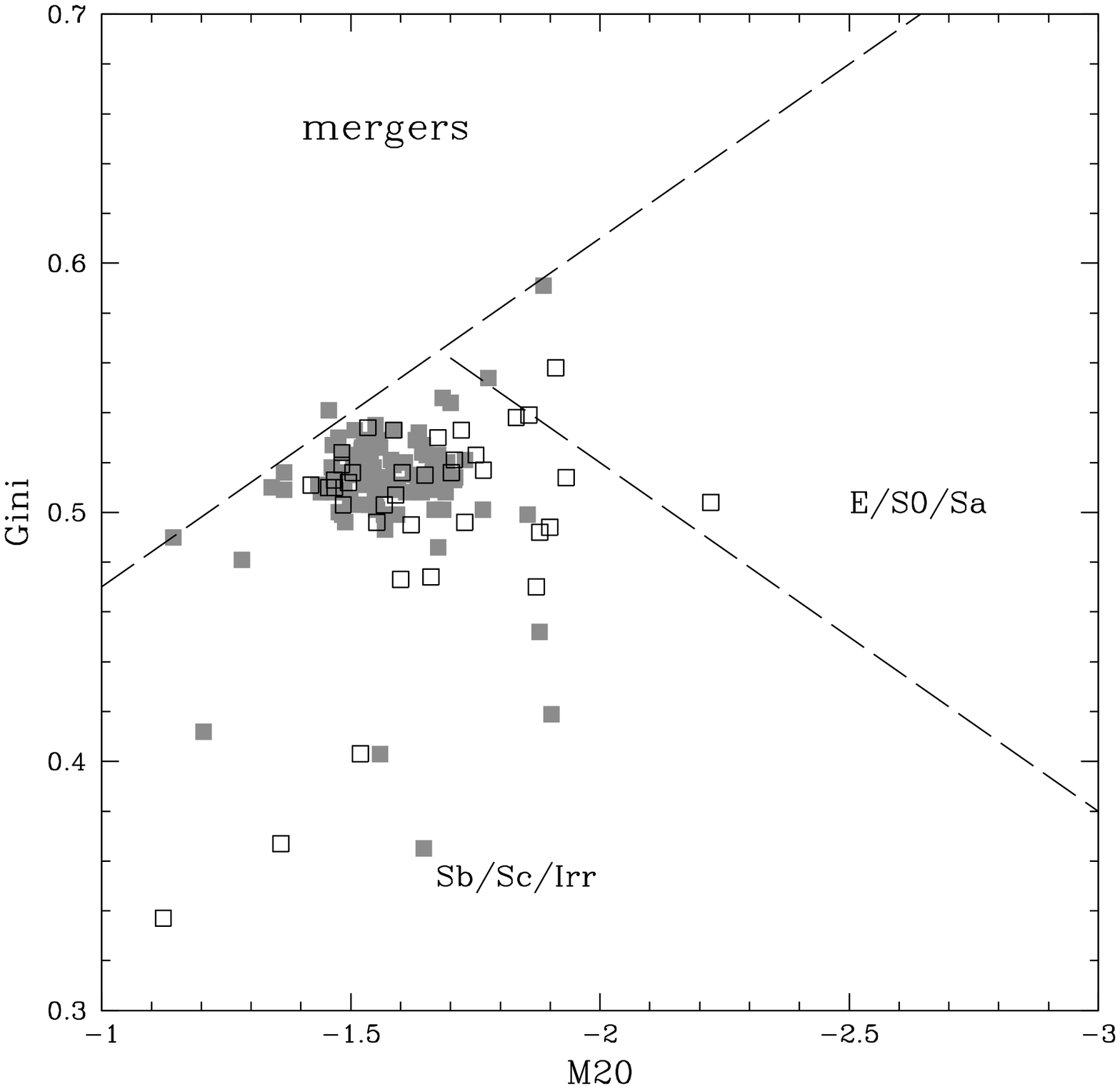}
	\caption{Gini vs. M20 coefficient (G) for obscured and unobscured (open and filled circles, respectively) AGNs (left panel) and obscured and unobscured (open and filled squares, respectively) QSOs (right panel). Dividing lines show regions of mergers, Sb/Sc/Irr and E/S0/Sa galaxy types taken from \citet{lotz08}.}
\label{gm20}
\end{figure*}

\subsection{Possible contamination by non-thermal light}

The presence of non-thermal light coming from the innermost parts of powerful QSOs can cause
biases in the measurements of some host galaxy parameters, such as colours, masses, etc.

\citet{pierce} describe the effect of AGN light on host galaxy optical ($U-B$) and UV-optical colours ($NUV-r$), 
as determined from X-ray selected AGN host galaxies at z$\sim$1.
These authors conclude that, in general, an active nucleus does not significantly affect the integrated colours of an
AGN host galaxy, even when near-UV colours are used.

\citet{hickox09} studied a sample of AGNs with $0.25<z<0.8$, selected in the radio, X-ray and IR bands from the
AGN and Galaxy Evolution Survey (AGES). These authors calculated the correction for AGN contamination in the $u-r$ host galaxy colours finding that the typical correction for nuclear contamination ranges from 0 to 0.3 mag.

In a similar way \citet{kauff07}, using a low-redshift sample of optically selected AGNs, found a lack of
colour contamination among the AGN host galaxies in their sample.

We also carried out a visual inspection of high resolution (0.03"/pixel) HST images (F814W filter) of our sample of AGNs finding that only $<$10\% of the total AGNs$+$QSOs present visible nuclear point source in their optical images \footnote{We have used the Cutouts Service from the NASA/IPAC Infrared Science Archive http://irsa.ipac.caltech.edu/data/COSMOS/index\_cutouts.html}.
Following the previous results, we estimate that the correction for AGN contamination on host galaxy colours is less than $\sim$0.3 mag. and so the results are not affected. The estimated maximum difference of 0.3 mag. corresponds approximately to the point size in Figure \ref{quiescent}.

\subsection{Morphological analysis}

In this section we analyse the morphology of the different AGN host galaxies according to non-parametric measures such as the Gini coefficient and the M20 parameter.
The Gini coefficient is defined as the absolute value of the difference between the integrated cumulative distribution of galaxy intensities and a uniform intensity distribution \citep{abraham}, i.e., quantifies to what extent light is equally distributed among the pixels in an image, regardless of their spatial location \citep{lotz}.
High Gini coefficient values mean that most of the light is coming from just a few pixels and low values implies that all pixels are more or less equally bright.

\begin{figure*}
\centering 
\includegraphics[width=0.47\textwidth]{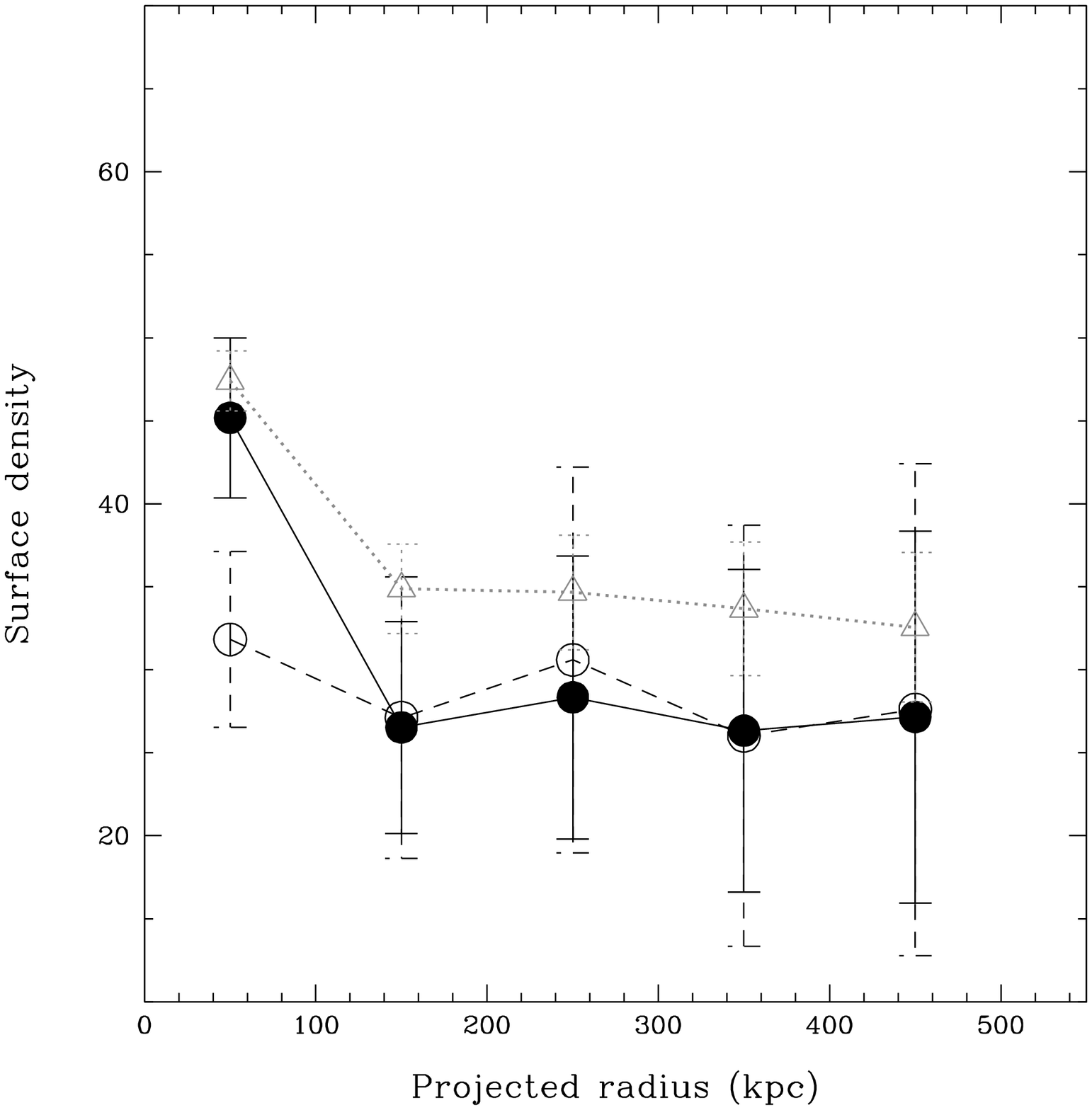}
\includegraphics[width=0.47\textwidth]{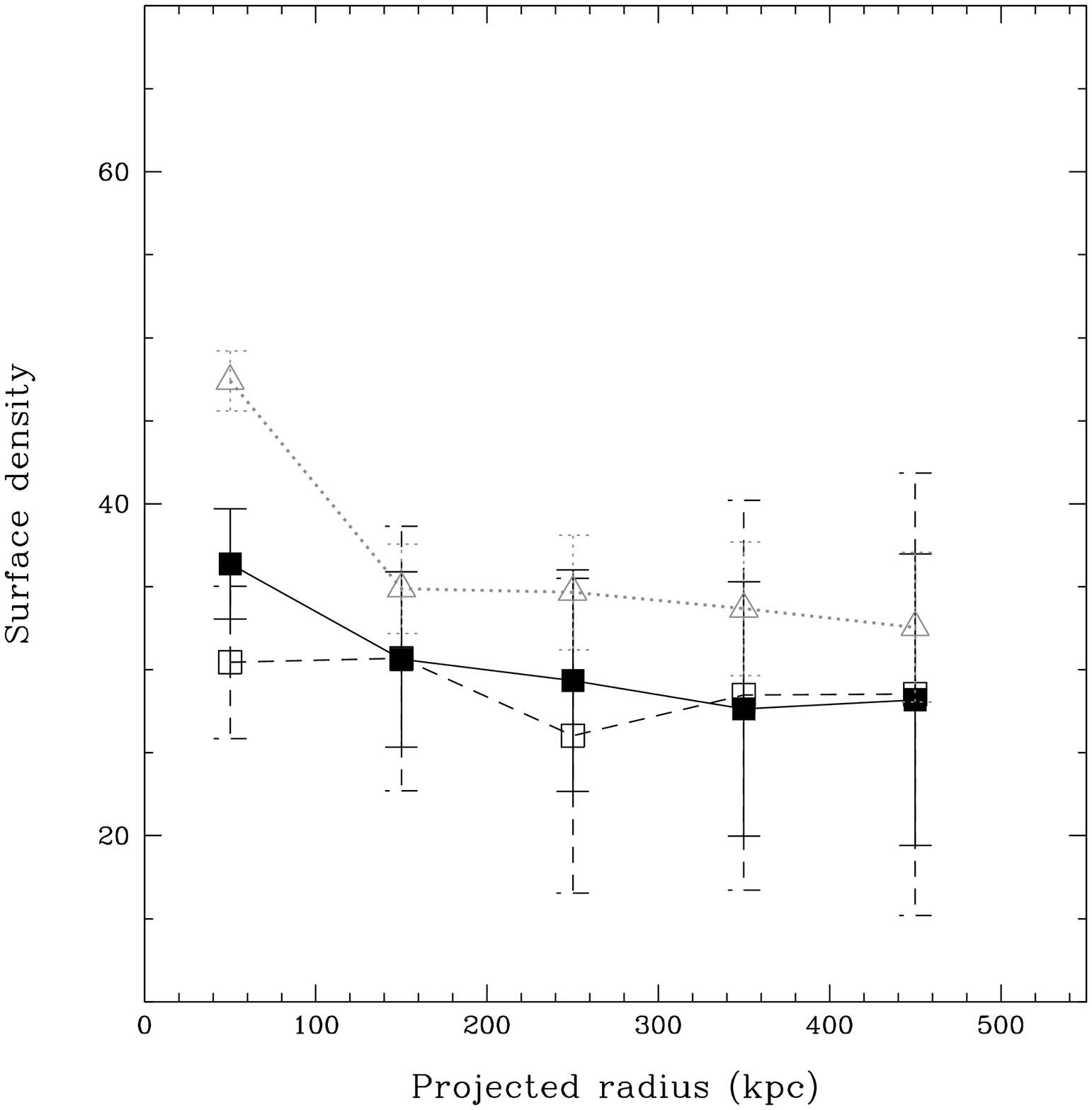}
	\caption{Projected radial density of tracer galaxies with $|z_{AGN}-z_{galaxy}|\leq0.2$ and $r<25.75$ around the selected obscured and unobscured (open and filled circles) AGNs (left panel) and for obscured and unobscured (open and filled squares) QSOs (right panel). Open triangles in each figure represent the corresponding values for SF galaxies. The error bars represent the standard deviation within each data bin, estimated using Poissonian errors.}
\label{pro}
\end{figure*}

The M20 parameter which was introduced by \citet{lotz}, measure the spatial extent of the brightest galaxy regions, computed by measuring the second-order moment of the brightest 20\% of the galaxy pixels, defined as the sum of the intensity of each pixel multiplied by the square of the distance from the centre of the galaxy for the brightest 20\% of the pixels in a galaxy. It is more sensitive to the presence of bright regions at large galactocentric distances and provides information of merger signatures such as multiple nuclei, tidal tails, bars, etc \citep{lotz,kong09,wang12}.

The morphological measures used in this work were obtained from the catalogue presented by \citet{cassata} which provides information of non-parametric diagnostics of galaxy structure using the Hubble Space Telescope Advanced Camera for Surveys (ACS) for 232022 galaxies up to F814W$=$25. 

We cross-correlated the spatial positions of AGNs previously selected from Chang et al. catalogue with those presented by \citet{cassata}, using a matching radius of 1 arcsec. We find the following matches: 75\% and 85\% for obscured and unobscured AGNs and 85\% and 94\% for obscured and unobscured QSOs.

In Figure \ref{gm20} we plot the M20 parameter vs. Gini coefficient for the sample of obscured and unobscured (open and filled circles) AGNs (left panel) and for obscured and unobscured (open and filled squares) QSOs (right panel).
The dashed line represent the criterion published by \citet{lotz08}, according to the following definitions, Mergers: $G >-0.14 M_{20}+0.33$, Early(E/S0/Sa): $G <-0.14 M_{20}+0.33$, and $G > 0.14 M_{20}+0.80$, Late(Sb/Sc/Ir): $G < -0.14 M_{20}+0.33$, and $G > 0.14 M_{20}+0.80$.

We find that only 4\% and 1\% of unobscured AGNs and QSOs, respectively are located in the region identified by mergers.

In the case of unobscured AGNs (filled circles, left panel), can be observed the presence of some objects in the area with low Gini indices ($G \lesssim 0.45$). A visual inspection of them indicated that they are objects with irregular morphology.

Some works show a  link between mergers and powerful obscured AGNs.
 \citet{glikman} from a study using the Hubble Space Telescope (HST) Wide Field Camera 3 (WFC3) near-infrared camera to image the host galaxies of a sample of 11 luminous, dust-reddened quasars at z$\sim$ 2 reveal that 8/11 of these quasars have actively merging hosts. 
Similar results were found by \citet{assef15,fan16} in a sample of powerful, heavily-obscured WISE-selected quasars at z $\sim$ 2 which show a large fraction of mergers in their host galaxies. On the contrary, other works do not show these trends.
In a sample of X-ray selected AGN at z$\sim$2, \citet{kocevski} found that AGNs do not exhibit a significant excess of distorted morphologies while a large fraction reside in late-type galaxies.
\citet{Schawinski} showed for a sample of X-ray selected AGNs with 1.5$<z<$3 selected from the Chandra Deep Field South that the majority (80\%) 
of these AGNs have low S\'ersic indices indicative of disk-dominated light profiles and undisturbed morphologies.

The lack of mergers signatures in our AGN and QSO samples observed in the M20 vs. Gini plane suggests that either secular evolution, disk
instability-driven  accretion,  minor  mergers, or  a  combination of the three, play a major role in triggering AGN activity at
these redshifts than previously thought \citep{Schawinski}.


\section{Galaxy density}

In this Section we analyse the environment of obscured and unobscured AGNs and QSOs.
According to the postulates of the unified model, the host galaxies of the different AGNs would exhibit similar properties and would be located in similar galactic environments.
In Figure \ref{pro} we show the projected radial density of tracer galaxies with $|z_{AGN}-z_{galaxy}|\leq0.2$ and $r<25.75$ (which is the limiting apparent magnitude obtain from the Subaru $r$ band number counts) around the obscured and unobscured AGNs (left panel) and QSOs (right panel) with $r_p<500$ kpc. 
We also include the sample of SF galaxies in this analysis as a comparison sample. Error bars were estimated using Poissonian errors.
For the AGN sample we find that in small scales ($<$100 kpc) unobscured objects show more neighbouring galaxies than the sample of obscured AGNs.
While SF galaxies have a density of neighbouring galaxies similar to the sample of unobscured AGNs.
In the case of the QSOs, we find that both obscured and unobscured samples have lower galaxy densities than that obtained for the SF galaxy sample. 
Despite uncertainties at smaller scales, unobscured QSOs also show, more neighbouring galaxies than the sample of obscured QSOs.
We have also performed tests using other redshift limits, changing the redshift difference cut to  $|z_{AGN}-z_{galaxy}|\leq0.1$ 
 and 0.15 does not change the results except that the number of tracer galaxies is reduced.

Environments of the different AGN samples show contradictory results at low and high redshifts in the literature.
\citet{alle11} analysed a sample of X-ray selected AGNs from the $XMM$-COSMOS survey with $z<$1.5 and found a difference in the environments between X-ray obscured and unobscured AGNs, who inhabit dark matter haloes with mass log($M_{obs}$[M$\odot$])$=13$ and log($M_{unobs}$[M$\odot$])$=13.3$. In a similar way, \citet{alle14} studied a sample of AGNs selected on the basis of a combined X-ray, optical spectra and photometric data from Chandra and $XMM$ in the COSMOS area. These authors found that unobscured AGNs at z$\sim$3 reside in 10 times more massive haloes compared to obscured AGNs.
\citet{gilli09} analysed a sample of AGNs selected from the $XMM$-COSMOS survey with median redshifts $\overline{z}\sim$ 1.
These authors studied several samples of AGNs, among them a sample of X-ray selected AGNs using a criterion according to the hardness ratio, a subsample with $HR >-$0.2 (obscured) and other with $HR<-$0.2 (unobscured). The spatial clustering of these two sample shows that obscured and unobscured AGNs inhabit similar environments.
\citet{melnyk} studied the large ($>$ 1 Mpc ) and small ($<$0.4 Mpc) scale environment of X-ray selected AGNs with z$\sim$1 in the XXL survey using the nearest neighbour distance method and the local galaxy density. No significant differences in the clustering of obscured (HR $>-$0.2) and unobscured (HR $<-$0.2) AGNs were found.


\section{Neighbour galaxy properties}

\subsection{Colour, stellar mass and SFR of neighbouring galaxies}

It is well known that the properties of galaxies depend on the environment in which they reside.
In this sense, one of the pioneering works on the effect of the environment on the properties of galaxies was that of \citet{dressler}.
This author studied the relation galaxy morphology–density in which, as the density of the environment
increases, an increasing elliptical and S0 population is found, along
with a corresponding decrease of those galaxies with a spiral morphology \citep{postman,cape11,alpa}. 
Other physical properties of galaxies that also depend on morphology such as colour \citep{tanaka,balogh}, and SFR \citep{grutz}. 
It also be expected that AGN properties are related on the environment, and it could be the environment that
triggers processes such as AGN accretion \citep{padilla,wang17}.

\begin{figure*}
\includegraphics[width=55mm]{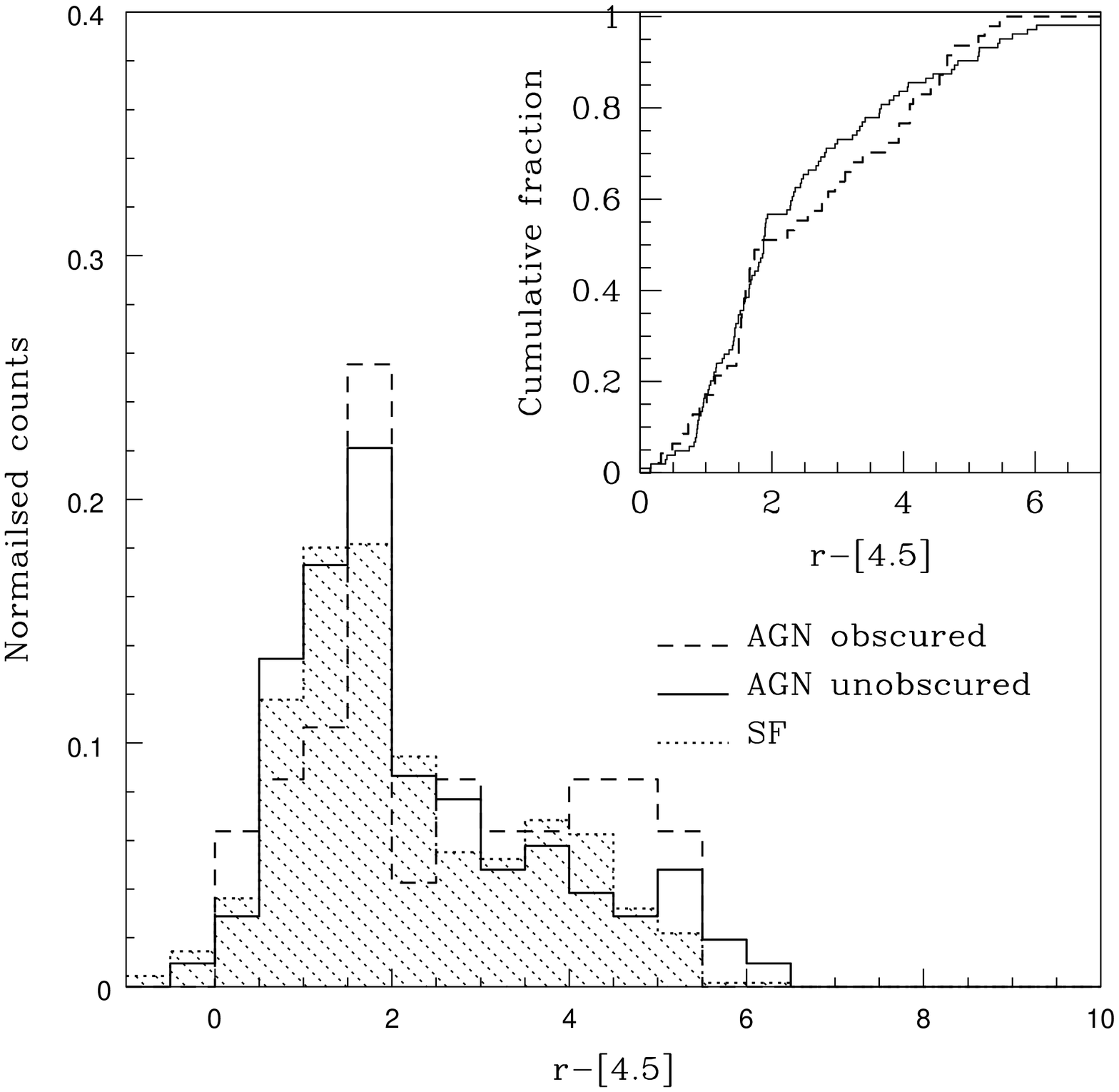}
\includegraphics[width=55mm]{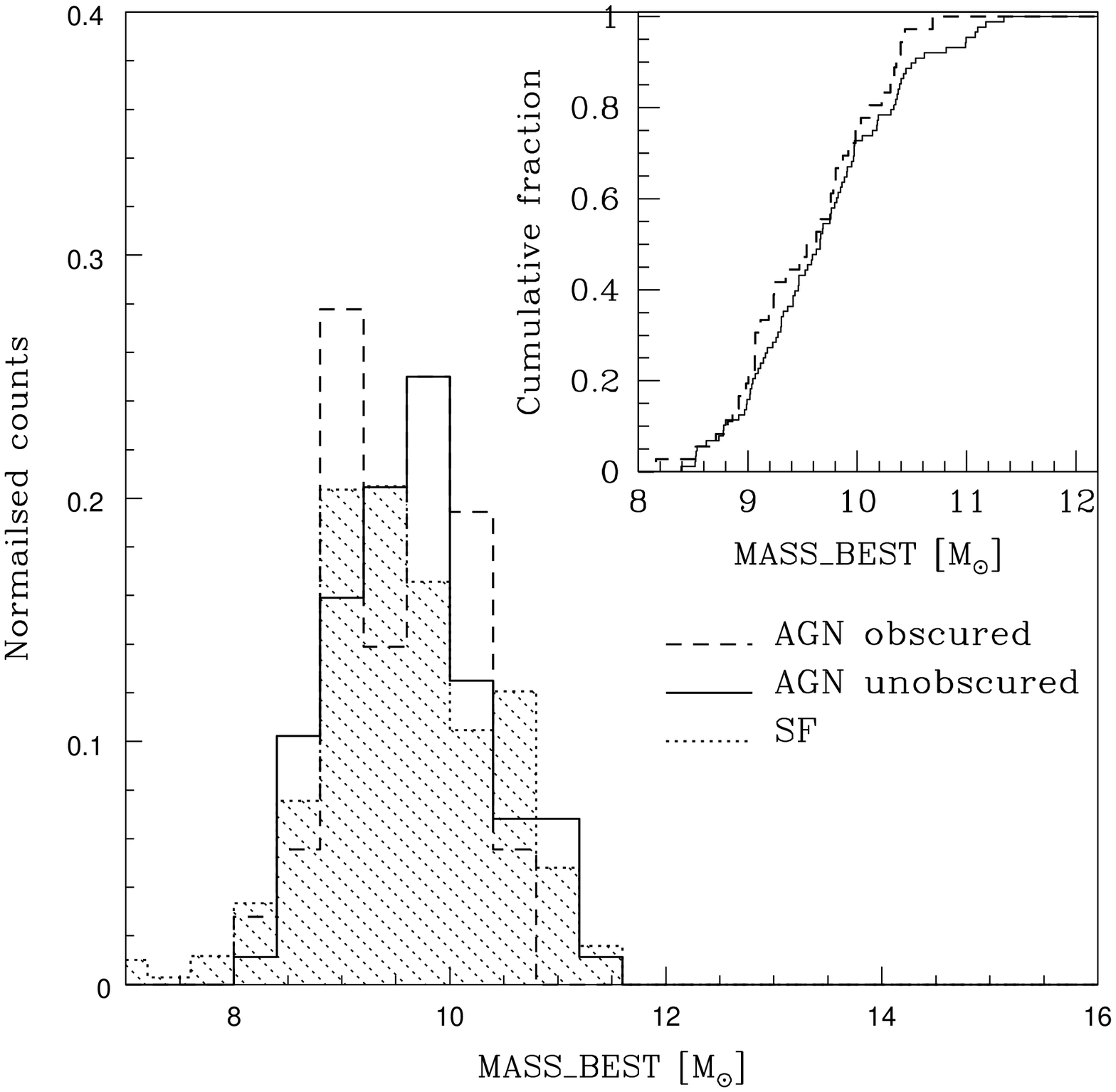}
\includegraphics[width=55mm]{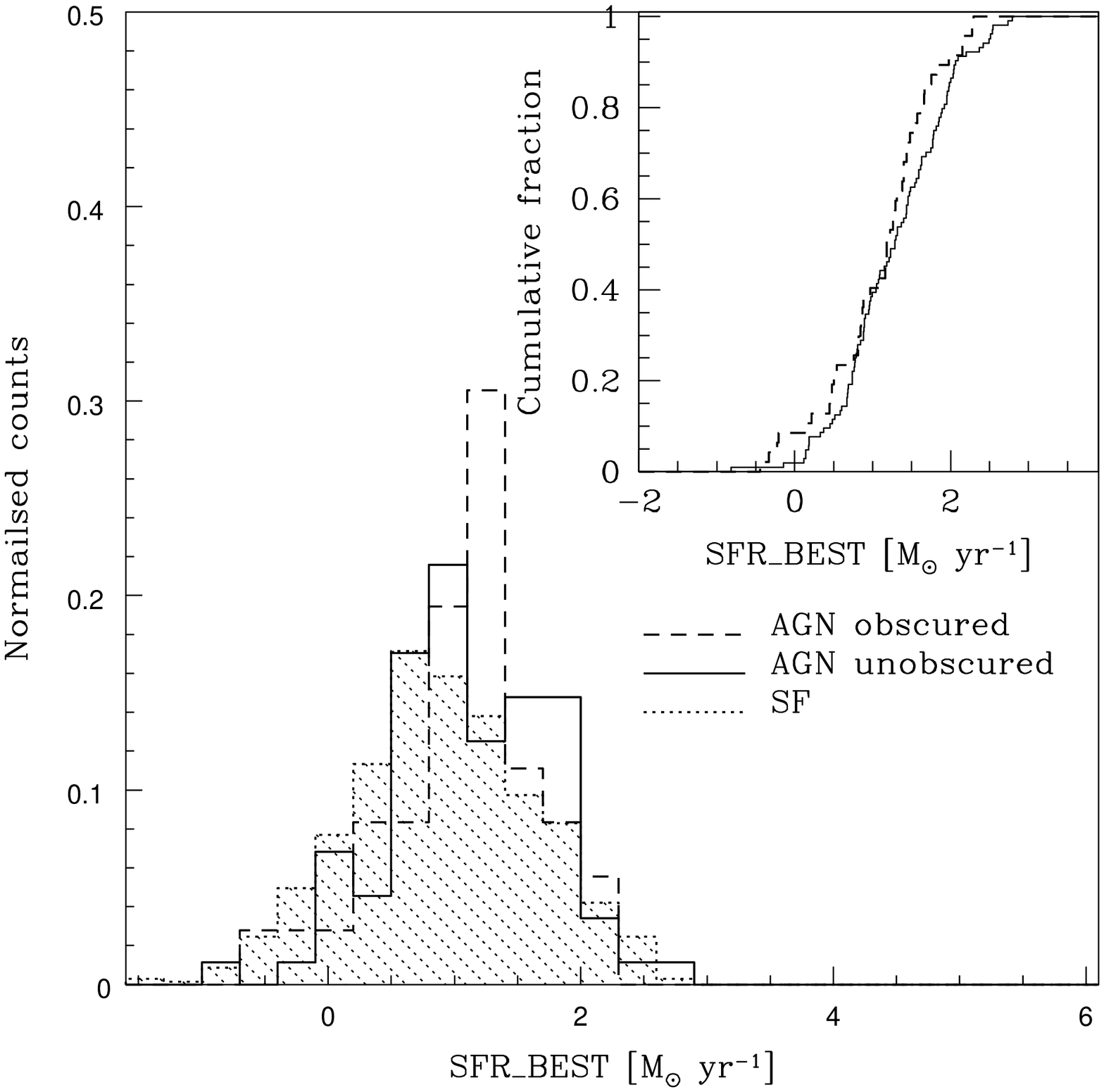}
	\caption{$r-[4.5]$ colour (left panel), stellar masses (middle panel), and SFR (right panel) distributions for tracer galaxies with $|z_{AGN}-z_{galaxy}|\leq0.2$ and $r<$25.75 within 100 kpc from obscured (dashed line histogram) and unobscured (solid line histogram) AGNs. Shaded histograms represent the corresponding values for tracer galaxies in the field of SF galaxies. The inset shows the cumulative fraction distributions.}
\label{100}
\end{figure*}

Here we study colours, masses, and SFRs of neighbour tracer galaxies located in the field of obscured and unobscured AGN and QSO samples.
In Figure \ref{100} we show the $r-[4.5]$ colour (left panel), stellar masses (MASS\_BEST, middle panel), and SFR (right panel) for tracer galaxies with $|z_{AGN}-z_{galaxy}|\leq0.2$ and $r<25.75$ within 100 kpc from obscured and unobscured AGN sample, which is the projected distance where we see differences in the environments. 
We have included the cumulative fraction distribution for the QSO and AGN samples (top right on each figure), in order to make clearer the differences between the different distributions.
We have also included the corresponding distributions of neighbouring galaxies of the sample of SF galaxies (shaded histogram).

We find that the colour distribution of tracer galaxies in the field of obscured AGN presents an excess of galaxies with red colours, compared to the samples of unobscured AGN and the sample of SF galaxies. We have used the GMM code in order to quantify the colour distribution.
For the sample of tracer galaxies in the field of obscured AGNs we find $\mu_{1}=1.4\pm0.1$ and $\mu_{2}=4.0\pm0.3$, while for unobscured AGNs we obtained $\mu_{1}=1.5\pm0.2$ and $\mu_{2}=4.0\pm0.6$. For the sample of tracer galaxies in the field of SF galaxies we find $\mu_{1}=1.40\pm0.05$ and $\mu_{2}=3.6\pm0.06$. 

The stellar mass distribution of tracer galaxies in the field of obscured, unobscured AGNs and the SF galaxy sample are similar (see Figure \ref{100}, middle panel). For the obscured and unobscured AGN sample we obtain $\overline{\mathrm{log}M_{obs}}=9.5\pm0.6$ ($M_{\odot}$) and $\overline{\mathrm{log}M_{unobs}}=9.6\pm0.7$ ($M_{\odot}$), while for the SF galaxies we find $\overline{\mathrm{log}M_{sf}}=9.5\pm0.8$ ($M_{\odot}$).

The SFR of tracer galaxies in the field of different AGNs and SF galaxies are slightly different. Comparing the values obtained for SF galaxies, tracer galaxies in the field of both obscured and unobscured AGNs have higher SFR values. For the obscured and unobscured AGN sample $\overline{\mathrm{log}(SFR_{obs}})=1.0\pm0.6$ and $\overline{\mathrm{log}(SFR_{unobs}})=1.1\pm0.6$. While for the SF galaxies we find $\overline{\mathrm{log}(SFR_{sf}})=0.9\pm1.0$. 

\begin{figure*}
\includegraphics[width=55mm]{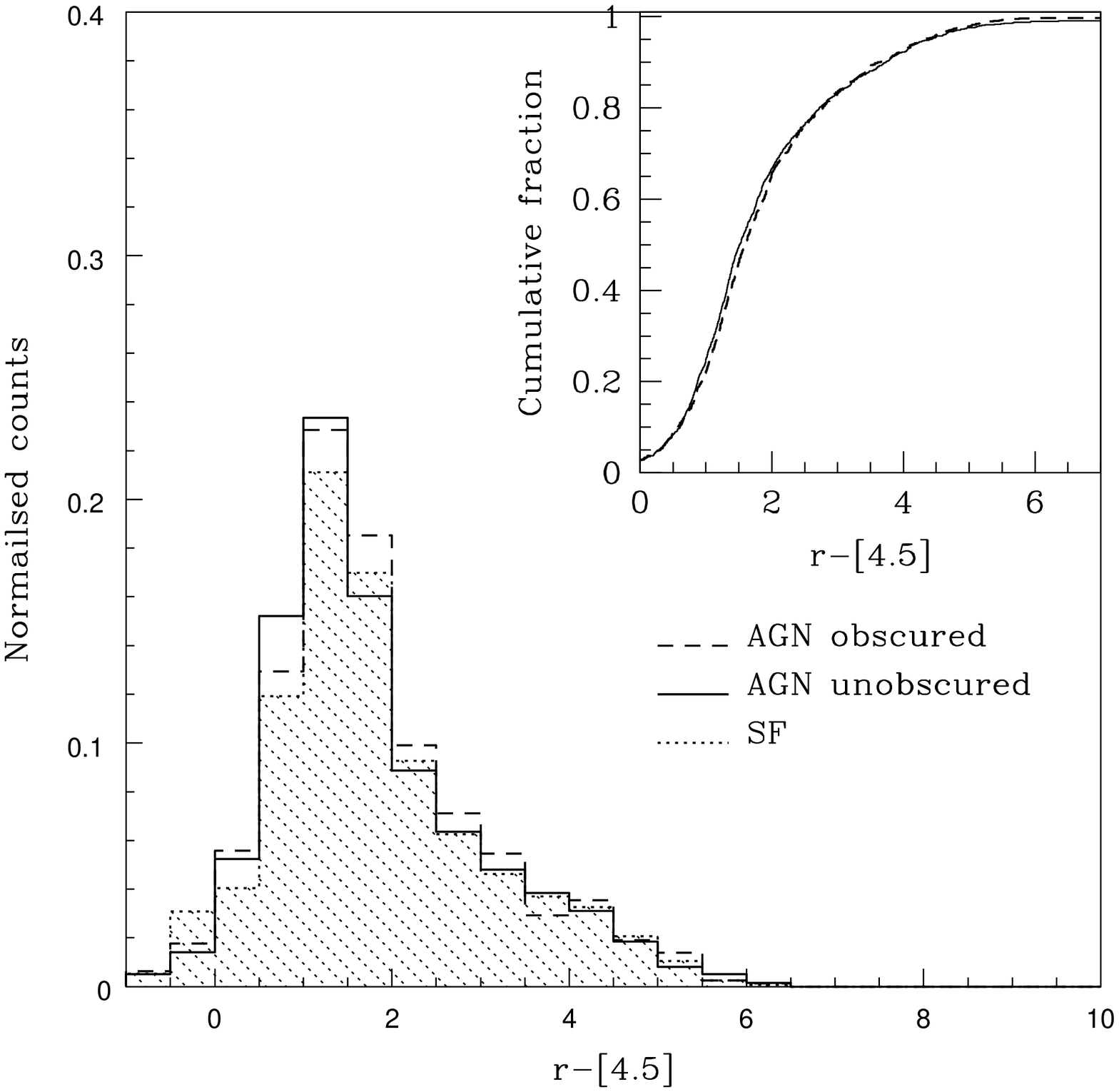}
\includegraphics[width=55mm]{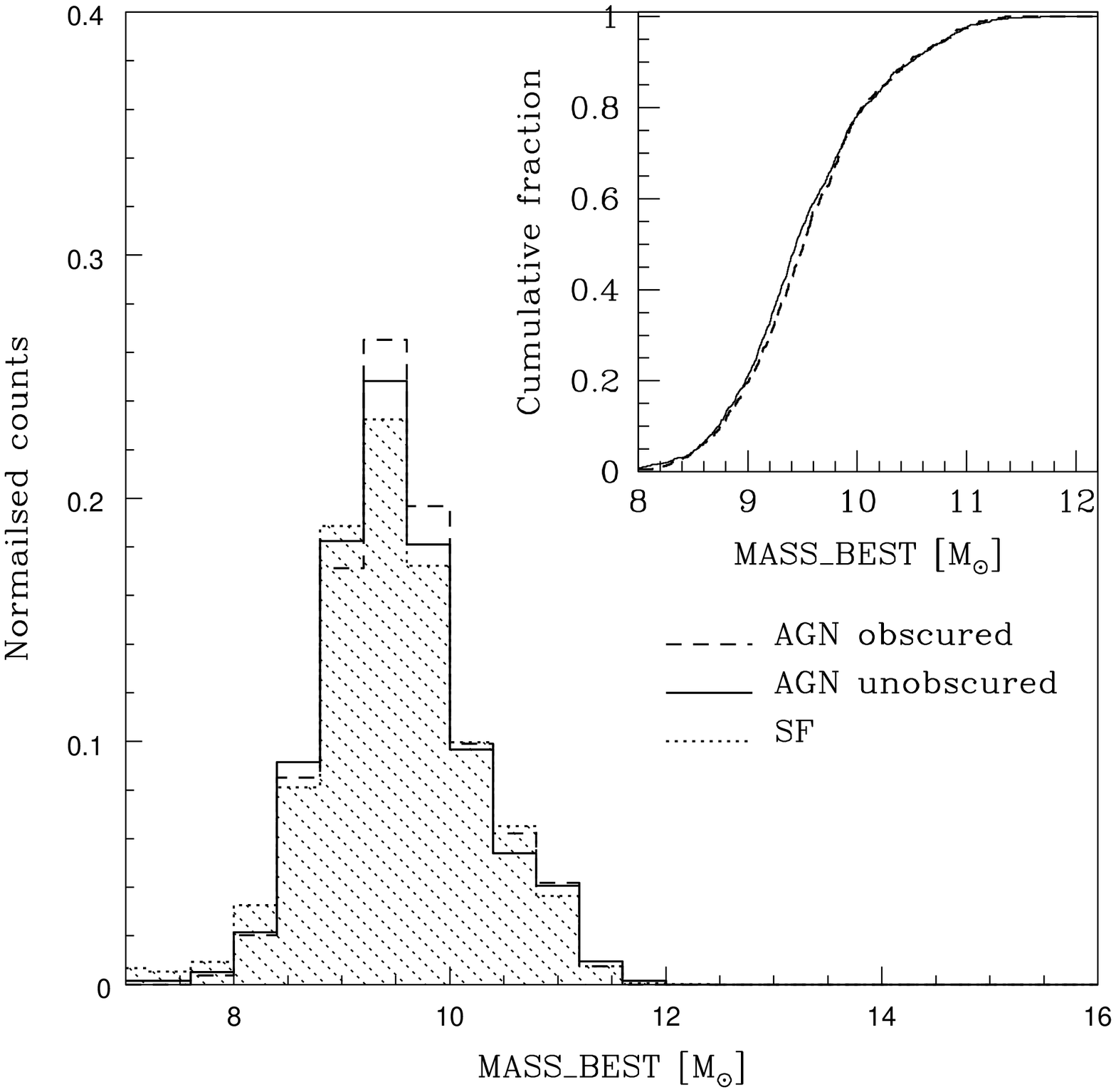}
\includegraphics[width=55mm]{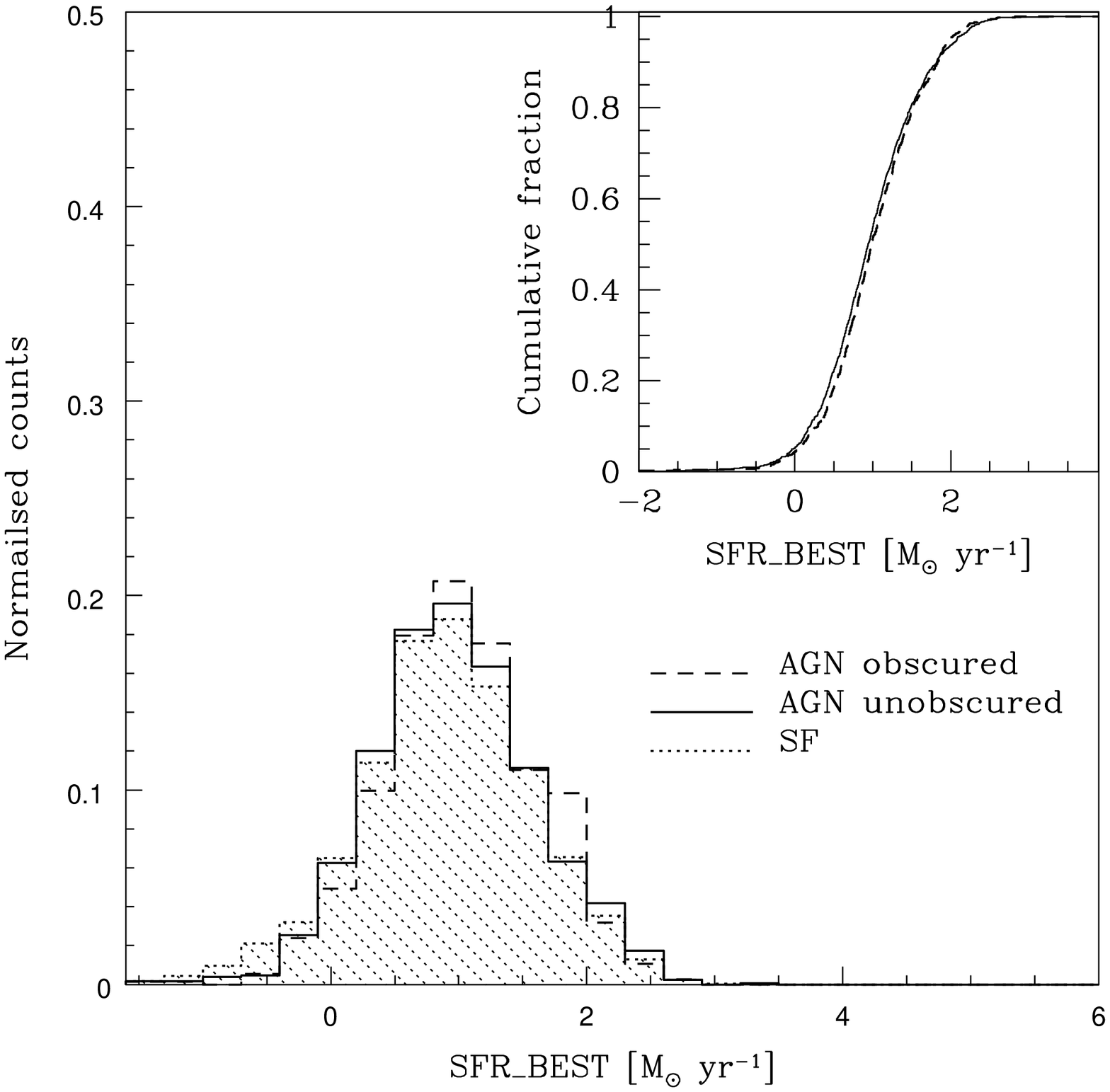}
	\caption{Same as Figure \ref{100} for a sample of tracer galaxies with 100 $<r_p<500$ kpc from the sample of AGNs}
\label{500}
\end{figure*}

In Figure \ref{500} we plot the same parameters as in Figure \ref{100} for a sample of tracer galaxies with 100 $<r_p<500$ kpc from the sample of AGNs.
As it can be seen the $r-[4.5]$, stellar mass and SFR of tracer galaxies in the field of obscured, unobscured and SF galaxies are similar. 

\begin{figure*}
\includegraphics[width=55mm]{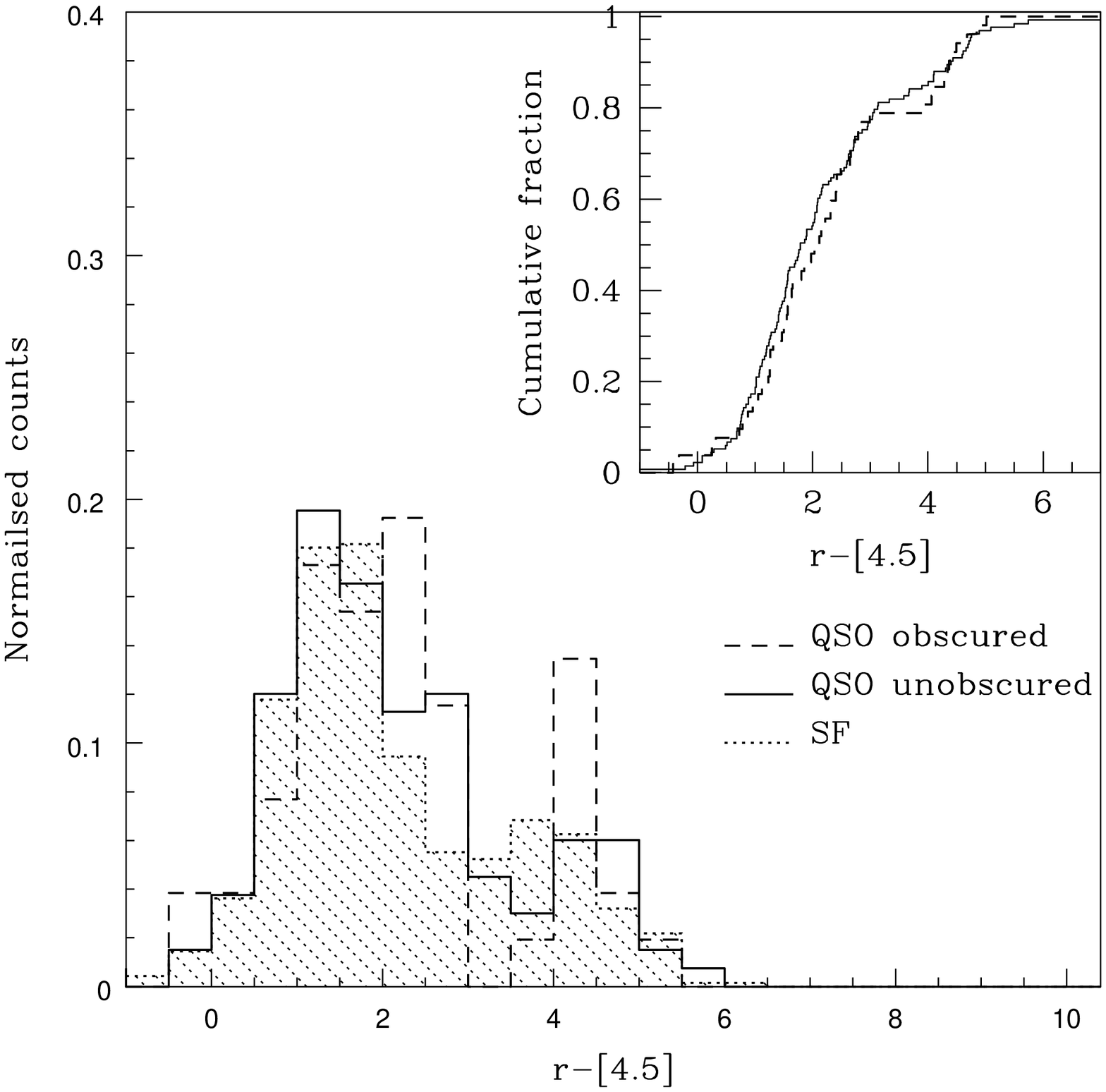}
\includegraphics[width=55mm]{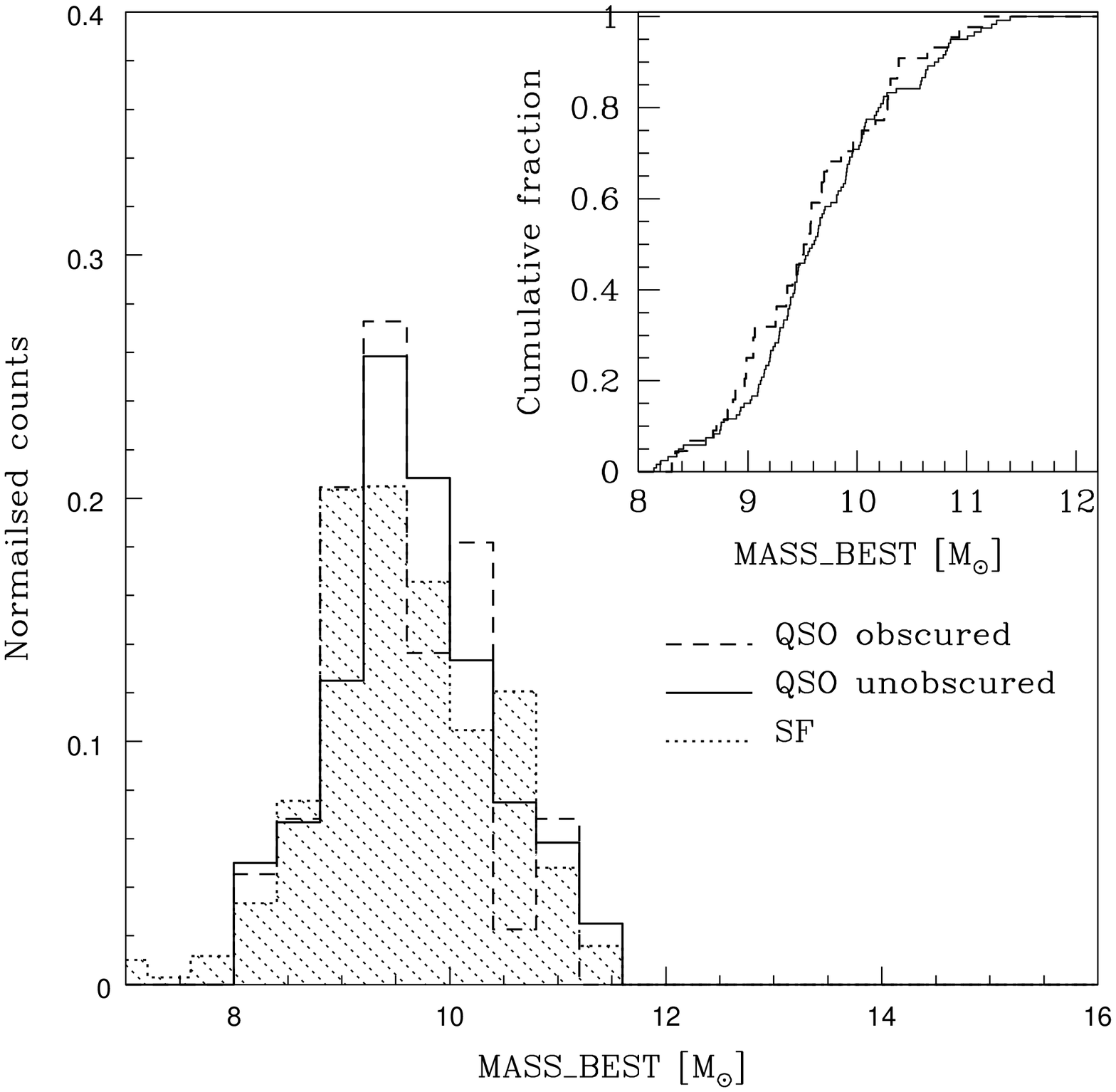}
\includegraphics[width=55mm]{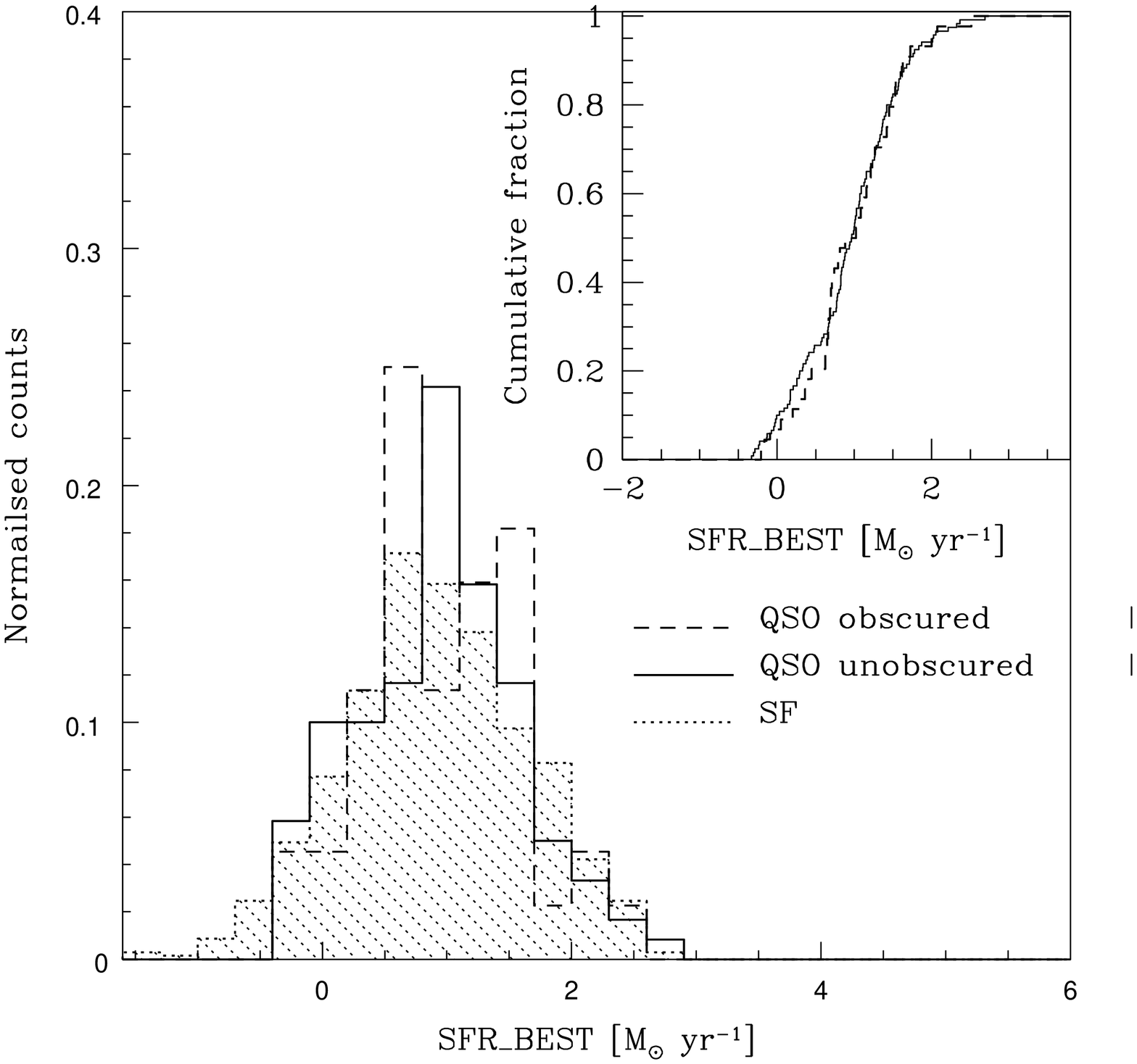}
	\caption{$r-[4.5]$ colour (left panel), stellar masses (middle panel), and SFR (right panel) distribution for tracer galaxies with $|z_{AGN}-z_{galaxy}|\leq0.2$ and $r<$25.75 within 100 kpc from obscured (dashed line histogram) and unobscured (solid line histogram) QSOs. Shaded histograms represent the corresponding values for tracer galaxies in the field of SF galaxies. The inset shows the cumulative fraction distributions.}
\label{100Q}
\end{figure*}

\begin{figure*}
\includegraphics[width=55mm]{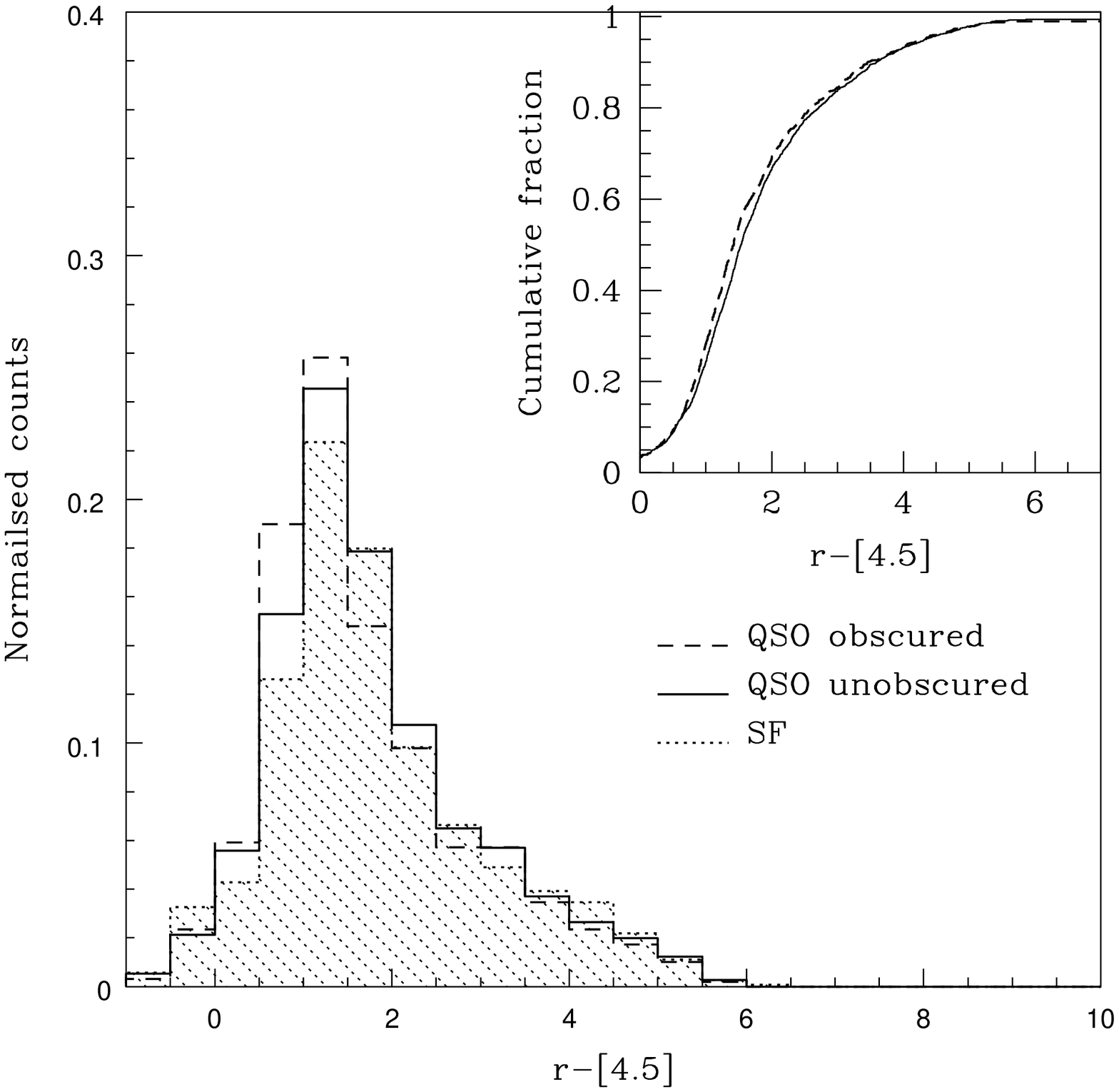}
\includegraphics[width=55mm]{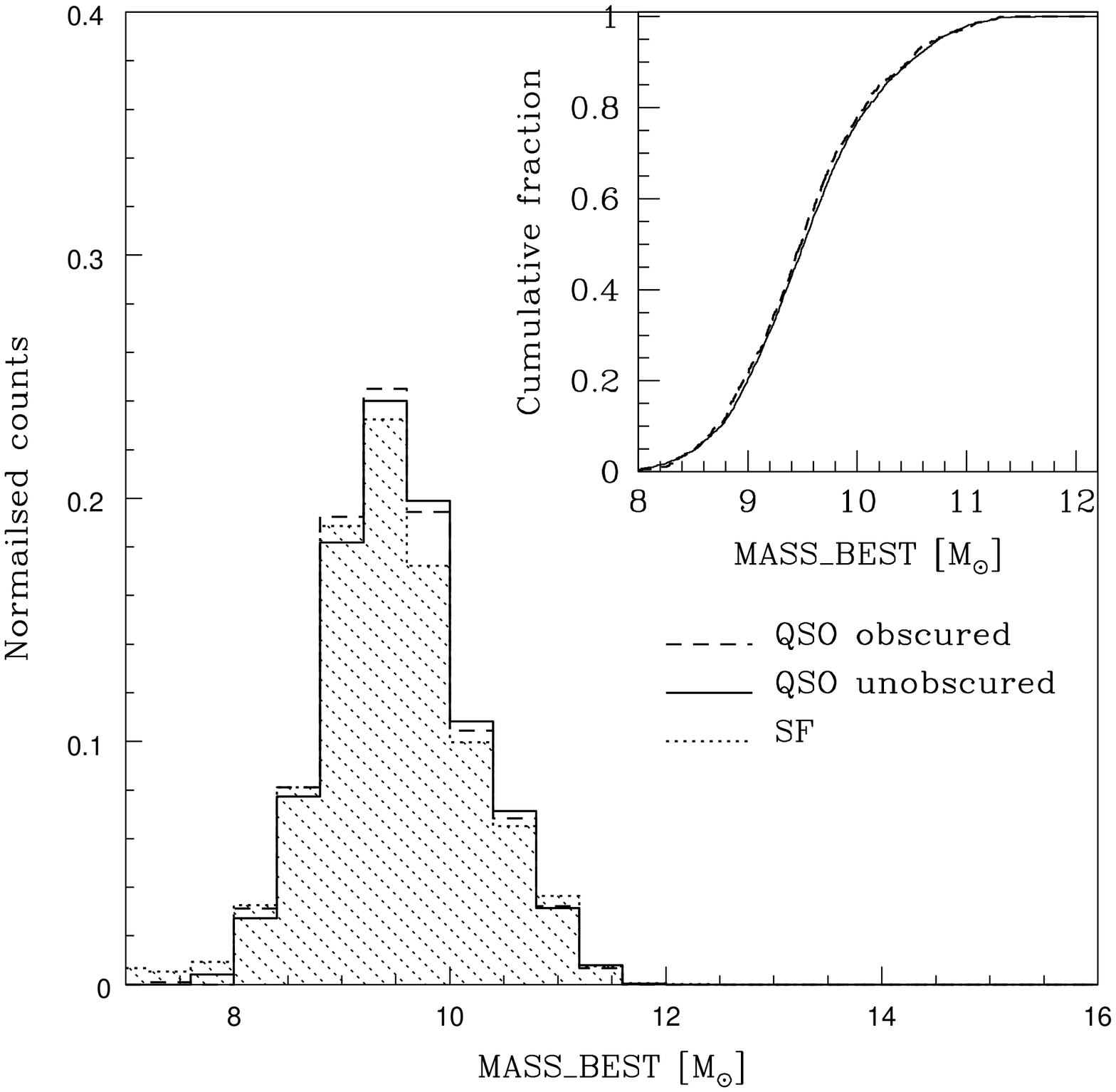}
\includegraphics[width=55mm]{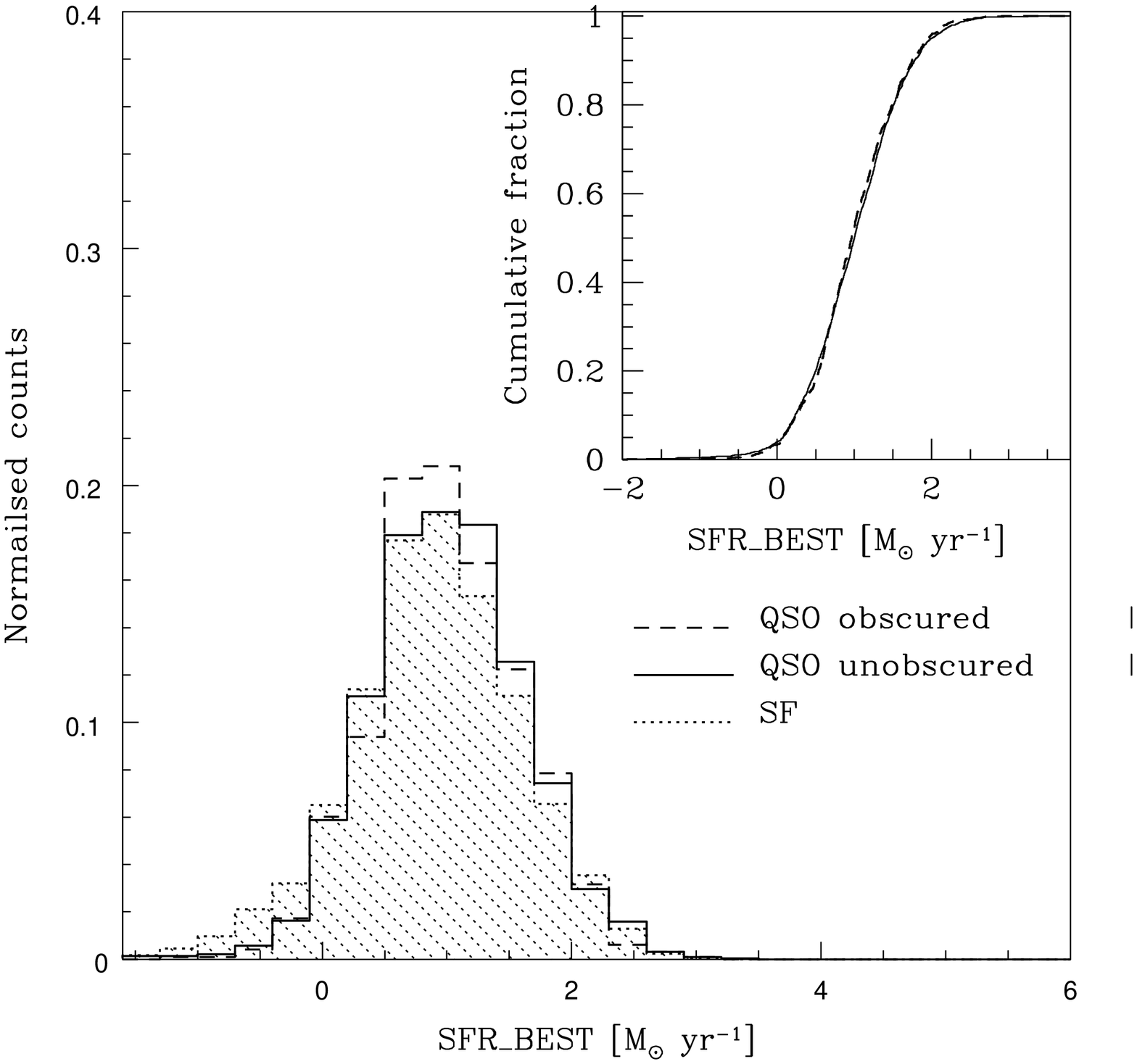}
	\caption{Same as Figure \ref{100Q} for a sample of tracer galaxies with 100 $<r_p<500$ kpc from the sample of QSOs}
\label{500Q}
\end{figure*}

In Figures \ref{100Q} and \ref{500Q} we performed a similar study for tracer galaxies in the field of QSOs.
In the left panel of Figure \ref{100Q} it can be seen that the $r-[4.5]$ colours of tracer galaxies in the field of obscured QSOs are redder in comparison with the unobscured and SF galaxy sample. Using the GMM code we obtained $\mu_{1}=1.7\pm0.1$ and $\mu_{2}=4.4\pm0.1$ for tracer galaxies in the field of obscured QSOs and $\mu_{1}=1.6\pm0.1$ and $\mu_{2}=4.4\pm0.3$ for the sample of unobscured QSOs. 

The stellar mass and the SFR of tracer galaxies do not present large differences between the different QSO and SF galaxy samples. For the stellar mass distribution we find: $\mu_{obs}=9.6\pm0.7$ and $\mu_{unobs}=9.7\pm0.7$, while for the SF galaxy sample we find $\mu_{sf}=9.5\pm0.8$. We find $\overline{\mathrm{log}(SFR_{obs}})=1.1\pm0.6$ ($M_{\odot}$ $yr^{-1}$) and $\overline{\mathrm{log}(SFR_{unobs}})=0.9\pm0.6$ ($M_{\odot}$ $yr^{-1}$) for SFR of tracer galaxies. For the SF galaxies we obtain $\overline{\mathrm{log}(SFR_{sf}})=0.9\pm1.0$ ($M_{\odot}$ $yr^{-1}$). 

We find similar results to those observed in the AGN sample for colours, stellar masses and SFRs of tracer galaxies with 100 $<r_p<500$ kpc of the different AGN types.


\subsection{Morphology of neighbouring galaxies}

\begin{figure*}
\includegraphics[width=150mm]{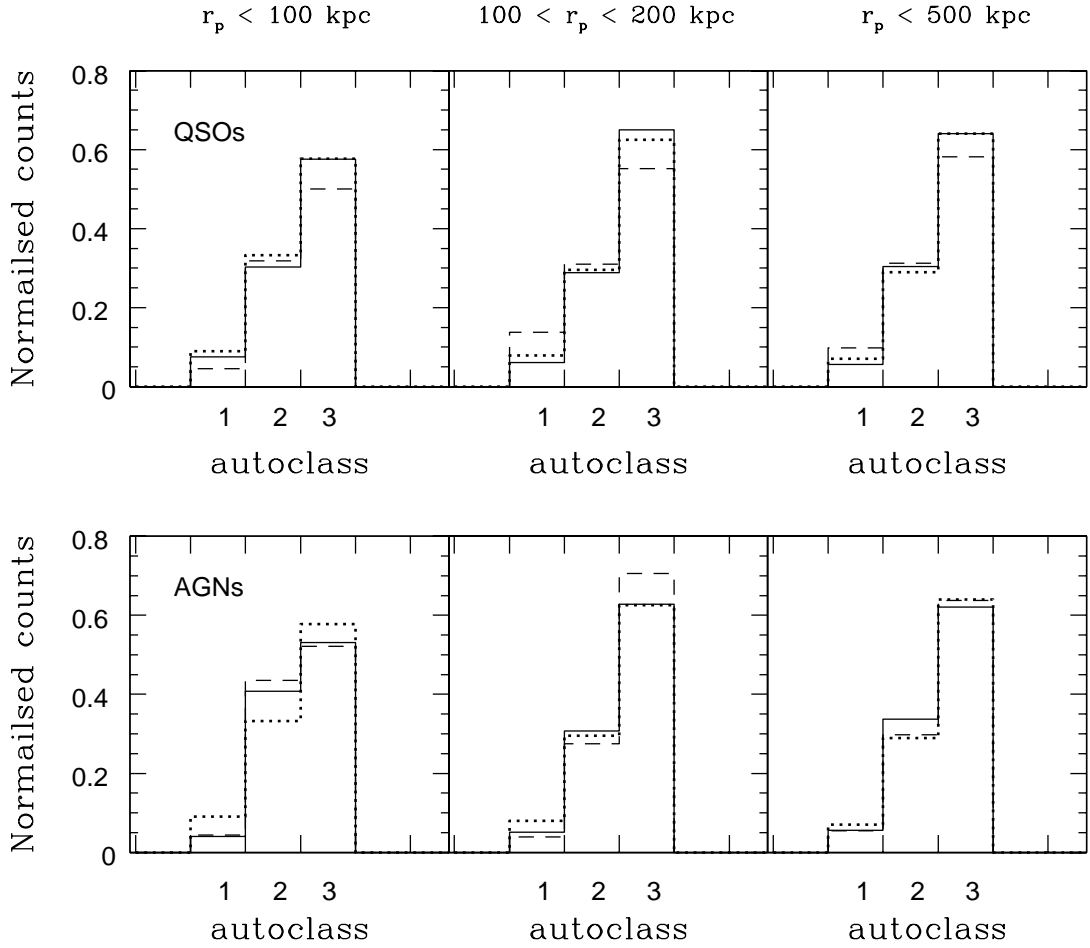}
	\caption{Morphology estimator AUTO\_CLASS distribution for tracer galaxies  with $|z_{AGN}-z_{galaxy}|\leq0.2$ and $r<25.75$ in the field of obscured (dashed lines) and unobscured (solid lines) QSOs (upper panel) and AGNs (lower panel). Dotted lines represent the corresponding values for tracer galaxies in the field of SF galaxies. AUTO\_CLASS values correspond to $1=$early-type; $2=$spiral; $3=$irregular.}
\label{vecinas}
\end{figure*}

We analyse in this section the morphology of the neighbouring tracer galaxies in the field of different AGN and QSO types.
We cross-matched our catalogue of tracer galaxies with $|z_{AGN}-z_{galaxy}|\leq0.2$ and $r<25.75$ in the field of AGN and QSO samples with the morphology catalogue presented by \citet{cassata}. 
We find that $\sim$ 40\% of the tracers have a counterpart in the catalogue of Cassata et al. 
The morphological parameters in Cassata et al. catalogue are combined to classify galaxies defined through the parameter AUTO\_CLASS as $1=$early-type; $2=$spiral; $3=$irregular.
We plot in Figure 14 the corresponding AUTO\_CLASS distribution in the field of obscured (dashed line histogram) and unobscured (solid line histogram) QSOs and AGNs. We included also the corresponding values for tracer galaxies in the field of SF galaxies (dotted lines). In all cases we divided the distribution for 3 different projected radius of $r_p<$100 kpc, 100$<r_p<$200 kpc and for $r_p<$500 kpc. 
At $r_p <$ 100 kpc, obscured and unobscured AGNs have a higher percentage of spiral tracer galaxies compared to the results obtained in the field of QSOs. For the QSO sample, we find a lower percentage of irregular galaxies in the field of obscured QSOs with $r_p<$100 kpc (see Figure \ref{vecinas}, left panel). 
For projected radius between 100$<r_p<$200 kpc, we find that obscured AGNs have more irregular tracer galaxies compared to the sample of obscured QSOs and SF galaxies.  
At $r_p <$ 500 kpc, both AGN, QSO and SF galaxies have similar percentage of early, spiral and irregular tracer galaxies.

\section{Summary and discussion}

In this paper we have studied the properties of host galaxies and their environment for a sample of obscured and unobscured X-ray selected AGNs and QSOs at high redshifts ($1.4\leqslant z \leqslant 2.5$) selected from Chang et al. catalogue in the COSMOS survey area. 

The main results can be summarised as follows:

\begin{itemize}

\item The optical-IR colour, $r -[4.5]$ of obscured and unobscured AGNs is bimodal. The colour separation threshold is consistent with the value found by \citet{hickox11,bornan17}, $r -[4.5]\sim 3$. The sample of obscured AGNs is divided into two parts: one consisting of host galaxies with predominantly red colours, and a small population with blue colours.
In the case of obscured QSOs, we find also that the colour distribution is bimodal, consisting in two similar population of objects with red and blue colours, while unobscured QSOs have predominantly blue hosts. 

\item The rest-frame colours of obscured and unobscured AGNs and QSOs observed in the $M_{NUV}-M_r$ vs. $M_r-M_J$ colour diagram show that most objects are located in the area defined by star-forming galaxies, rather than quiescent galaxies.
We also find bimodalities in $M_{NUV}-M_r$ colour distribution in unobscured AGN and QSO samples as well as in obscured QSOs.

\item The stellar mass vs. size relation for AGN and QSO hosts shows that in general, these objects are more compact and massive in comparison to a control sample SF galaxies. The stellar mass distribution of both AGN and QSO hosts are consistent with massive galaxies compared to the sample of SF galaxies. The stellar mass distribution of obscured and unobscured AGNs are similar, while unobscured QSOs are found more massive than obscured ones by $\Delta\overline{\mathrm{log}M}=0.3$ dex ($M_{\odot}$).

\item From a morphology analysis using non-parametric measures such as the Gini coefficient and the M20 parameter, we obtained that most of the AGNs and QSOs are located in regions belonging to late-type galaxies or irregulars. Only a small percentage are related to early-type galaxies and mergers.

\item The projected radial density of tracer galaxies around the sample of AGNs and QSOs show that both obscured and unobscured AGNs and QSOs inhabit similar galaxy environment. Only at small scales ($<$100 kpc) unobscured AGNs show more neighbouring galaxies than the sample of obscured AGNs. Unobscured AGNs show a similar radial profile as the sample of SF galaxies. For QSOs we find that both obscured and unobscured samples have lower galaxy densities than that
obtained for the SF sample. Despite errors at smaller scales, unobscured QSOs show also, more neighbouring galaxies than the sample of obscured QSOs. Our results are consistent with those previous obtained by \citet{alle14} in the sense that Type I AGNs reside in high density regions at high redshifts which may have evolved earlier. This contrasts with \citet{melnyk} findings who obtained similar environment for the different AGN types.

\item We find that the colour distribution of tracer galaxies with $r_p<100$ kpc around the field of obscured AGN presents an excess of galaxies with red colours, compared to the samples of unobscured AGN and the sample of SF galaxies.
The stellar mass distribution of tracer galaxies with $r_p<100$ kpc in the field of obscured, unobscured AGNs and SF galaxy sample are similar. The SFR of tracer galaxies with $r_p<100$ kpc around obscured and unobscured AGNs present similar distributions and have values larger than the SFR of tracer galaxies in the surroundings of SF galaxies.

For the case of QSOs, the stellar mass and the SFR of tracer galaxies do not present large differences.

\item From an analysis of the morphology of tracer galaxies in the field of different AGN types we find that: obscured and unobscured AGNs have a higher percentage of spiral tracer galaxies compared to the results obtained in the surroundings of both obscured and unobscured QSOs at within $r_p <$ 100 kpc. In the case of QSOs, we find a lower percentage of irregular galaxies in the field of obscured QSOs with $r_p<$100 kpc. Unobscured QSOs and SF galaxies present equal number of irregular tracer galaxies.

\end{itemize}

The differences of host galaxy properties such as UV/optical/IR colours and masses, together with the differences found in the projected galaxy density at small scales ($r_p<$100 kpc) and neighbouring galaxy properties, favour an evolutionary scenario rather than a strict unified model.

Although the unified model may be valid in some cases (for example at low redshifts), large-scale evolutionary processes undergone by a galaxy can cause its nucleus to become either obscured or unobscured during its lifetime.
Since the AGN existence is short ($\lesssim$ 0.001 Gigayear, \citealt{schawi}) compared to the merger stage ($\lesssim$ 1 Gigayear, \citealt{conse06}), possibly an important part of the galaxies repeatedly experienced AGN activity throughout its history due to gas infall to the central engine by disc instabilities or secular processes. Since the objects studied in this work are at high redshifts and the X-ray luminosity of these AGNs is 10-100 times greater than those at $z<1$, unobscured AGN or QSO phase might probably correspond to the end of an obscured phase where the growing BHs can produce highly enough accretion luminosity to sweep the surrounding material.







\section{Acknowledgments}
We sincerely thank the anonymous referee for useful suggestions, which improved the the quality of this manuscript.
Based on data products from observations made with ESO Telescopes at the La
Silla Paranal Observatory under ESO programme ID 179.A-2005 and on
data products produced by TERAPIX and the Cambridge Astronomy Survey
Unit on behalf of the UltraVISTA consortium.
Based on data obtained with the European Southern Observatory Very Large Telescope, Paranal, Chile, under Large Programs 175.A-0839 (zCOSMOS), 179.A-2005 (UltraVista) and 185.A-0791 (VUDS).
This work was partially supported by the Consejo Nacional de Investigaciones Cient\'{\i}ficas y T\'ecnicas (CONICET) and the Secretar\'ia de Ciencia y Tecnolog\'ia de la Universidad de C\'ordoba (SeCyT).


{}

\end{document}